\documentclass[nonacm]{acmart}

\AtBeginDocument{%
  }

\setcopyright{none}
\acmConference[arXiv]{arXiv preprint}{2025}{}

\begin{document}

\title[The Evaluation for Usability Methods of Unmanned Surface Vehicles]{Are Current Usability Methods Viable for Unmanned Surface Vehicles? Insights from a Multiple Case Study Approach to Human-Robot Interaction}


\author{Zitian Peng}
\email{z.peng7@liverpool.ac.uk}
\orcid{0009-0002-7496-6847}
\affiliation{%
  \institution{University of Liverpool}
  \city{Liverpool}
  \country{United Kingdom}
}
\affiliation{%
  \institution{Xi’an Jiaotong-Liverpool University}  
  \city{SuZhou}                                   
  \country{China}                     
}

\author{Shiyao Zhang}
\email{Shiyao.Zhang14@student.xjtlu.edu.cn}
\orcid{0009-0000-1591-3759}
\affiliation{%
  \institution{University of Liverpool}
  \city{Liverpool}
  \country{United Kingdom}
}
\affiliation{%
  \institution{Xi’an Jiaotong-Liverpool University}  
  \city{SuZhou}                                   
  \country{China}                     
}

\author{Shanliang Yao}
\email{Shanliang.Yao@ycit.edu.cn}
\orcid{0000-0001-7596-3598}
\affiliation{%
  \institution{Yancheng Institute of Technology}
  \city{Yancheng}
  \country{China}
}

\author{Xiaohui Zhu}
\email{Xiaohui.Zhu@xjtlu.edu.cn}
\orcid{0000-0001-7695-4538}
\affiliation{%
  \institution{University of Liverpool}
  \city{Liverpool}
  \country{United Kingdom}
}
\affiliation{%
  \institution{Xi’an Jiaotong-Liverpool University}  
  \city{SuZhou}                                   
  \country{China}                     
}

\author{Mengjie Huang}
\email{Mengjie.Huang@xjtlu.edu.cn}
\orcid{0000-0001-8163-8679}
\affiliation{%
  \institution{University of Liverpool}
  \city{Liverpool}
  \country{United Kingdom}
}
\affiliation{%
  \institution{Xi’an Jiaotong-Liverpool University}  
  \city{SuZhou}                                   
  \country{China}                     
}

\author{Prudence Wong}
\email{P.Wong@liverpool.ac.uk}
\orcid{0000-0001-7935-7245}
\affiliation{%
  \institution{University of Liverpool}
  \city{Liverpool}
  \country{United Kingdom}
}

\author{Yue Yong}
\email{Yong.Yue@xjtlu.edu.cn}
\orcid{0000-0001-7695-4538}
\affiliation{%
  \institution{University of Liverpool}
  \city{Liverpool}
  \country{United Kingdom}
}
\affiliation{%
  \institution{Xi’an Jiaotong-Liverpool University}  
  \city{SuZhou}                                   
  \country{China}                     
}

\renewcommand{\shortauthors}{Zitian et al.}

\begin{abstract}

Unmanned Surface Vehicles (USVs) are increasingly utilised for diverse applications, ranging from environmental monitoring to security patrols.  While USV technology is progressing, it remains clear that full autonomy is not achievable in all scenarios, and remote human intervention is still crucial, particularly in dynamic or complex environments. This continued reliance on human intervention highlights a range of Human-Robot Interaction (HRI) challenges that remain unresolved. Compared to the extensive body of HRI research in domains such as unmanned aerial vehicles and autonomous vehicles, HRI considerations specific to USVs remain significantly underexplored. Addressing this gap, our study investigates real-world usability challenges in USV operation through in-depth interviews with 9 engineers and users, supported by field observations. We focus especially on the difficulties beginner operators encounter and their coping strategies. 
Our findings reveal existing usability issues, mental models, and adaptation strategies of beginners that inform future user-centered design of USV systems, contributing new insights to the emerging field of maritime HRI.
Based on these findings, we argue that current USV systems are poorly suited for beginner operation in dynamic inland and offshore environments, where operators must make timely decisions under uncertainty, manage complex spatial awareness, and adapt to changing environmental conditions.
Furthermore, we identify key operational patterns in three representative use cases-harmful algal bloom detection, underwater concealed pipe inspection and post-construction hydrographic survey, and summarise key interaction constraints that should inform future maritime HRI design efforts.
\end{abstract}




\maketitle

\section{Introduction}
Unmanned Surface Vehicles (USVs), equipped with percepting sensors and automating algorithms, aim to autonomously navigate water surfaces without continuous human supervision. They are often deployed in challenging environments such as narrow rivers, vast lakes, deep oceans, or disaster-stricken areas, performing a variety of tasks, including detection operations \cite{liang2022multi,wu2023cooperative}, rescue missions \cite{jorge2019survey,zhang2022event}, and intelligent transportation \cite{cho2020efficient, liu2022intelligent}. 


Although USVs ultimately aim for full automation driven by machine intelligence, current technological limitations—especially in complex navigation environments—still require human intervention. Consequently, Human-Robot Interaction (HRI) research in unmanned systems, including USVs, Autonomous Vehicles (AVs), and Unmanned Aerial Vehicles (UAVs), has gained increasing attention due to the growing need for human oversight in challenging scenarios.
A key focus of these studies is evaluating usability, consistency, and user experience across different operational scenarios, as they critically influence operational efficiency, safety, and trust in human-robot teams. In particular, interaction interfaces for AVs and UAVs have been extensively explored to enhance interaction efficiency and reduce user cognitive load. For instance, Agrawal et al. \cite{Agrawal} 
addressed situational awareness challenges in emergency UAV operations by engaging domain experts in a participatory design process. Their work specifically tackled cognitive overload by identifying common attention failures and integrating mitigation strategies early in the interface design process.
Huibin et al. \cite{jin2019usability} 
evaluated the UAV ground control interface through questionnaires, interviews, and eye-tracking experiments. Their results highlighted usability challenges such as memory burden and poor user guidance, and demonstrated that an improved interface significantly enhanced task efficiency and user focus. 
Similarly, Hoang et al. \cite{Hoang} 
explored human-swarm interaction in multi-UAV emergency scenarios through professional user studies. Their studies highlighted interface-related cognitive challenges—such as managing multiple video feeds, coordinating attention across screens, and ensuring team communication—pointing to the need for interaction designs that mitigate information overload in real-world deployments.
These works highlight the increasing complexity of unmanned systems and underscore the importance of human-in-the-loop design to ensure operational safety and effectiveness.

Recent research increasingly highlights the importance of contextualised use case analysis and the explicit identification of user roles in the design of effective HRI systems. These approaches are particularly valuable in complex operational domains, where human tasks and decision-making vary significantly across scenarios and stakeholders. For instance, Ljungblad et al. \cite{ljungblad2021matters} analysed the practices of professional drone pilots and argued that many HRI concepts lack grounding in real-world operational contexts, proposing design implications rooted in actual workflows. Tener and Lanir \cite{tener2022driving} developed a teleoperation framework for autonomous vehicles through expert interviews and field observations, uncovering six context-specific challenges that shape interface design. Dong et al. \cite{Dong} explored holistic interface design by involving both drivers and external road users in participatory seminars, illustrating how use context and stakeholder perspectives jointly inform HRI requirements. Lingam et al. \cite{lingam2024challenges} further emphasised the importance of clearly defining human roles—such as operator, recipient, and bystander—during UAV deployments, showing how different roles give rise to distinct interaction needs and safety considerations.
Together, these studies demonstrate that without careful attention to domain-specific scenarios and human role differentiation, HRI systems risk failing to address the nuanced demands of real-world deployment.

While HRI research in domains such as UAVs and AVs has explored interaction design, operator workload, and interface usability in rich, real-world contexts, similar investigations remain scarce in the context of USVs. This is despite the fact that USVs share key operational characteristics with other unmanned systems—such as remote control, semi-autonomy, and mission-critical deployments—which suggests that some HRI design insights could be transferable. 
However, USVs also present unique interaction challenges, stemming from the interplay between aquatic environments, communication constraints, and the inherent complexity of vessel dynamics.
These specific constraints demand tailored HRI solutions rather than direct adoption of UAV or AV paradigms \cite{Ferreira}.

At the same time, the operational use of USVs is rapidly expanding into domains such as environmental monitoring, hydrological surveys, and infrastructure inspection—making the absence of contextualised HRI research particularly pressing. While UAV and AV studies have demonstrated the value of scenario-based user research in uncovering domain-specific interaction demands—especially those faced by novice operators and mixed-experience teams \cite{Acharya}—we still know little about how human users engage with USV systems in the field. Questions remain around how operators build mental models, what usability issues emerge across different roles, and how these factors affect task execution in realistic use cases.

Motivated by these gaps, our work adopts an empirical approach commonly used in UAV and AV HRI research, interviewing domain experts and observing real-world operations—to investigate USV usability, which is currently underexplored. We conducted several in-depth interviews with 9 key stakeholders from five companies involved in USV development and operations. These stakeholders include USV control engineers who oversee field operations, solution architects responsible for creating algorithms and control software, and end-users such as university researchers. To complement the interviews, we observed multiple field operations where engineers controlled USVs to accomplish diverse tasks requested by clients. These observations not only allowed us to capture subtle usability issues that participants might not have articulated, but also helped us identify typical use cases and operational challenges. Our goal was to evaluate the usability of existing USV systems, to understand the mental models and adaptation strategies of novice users, and to uncover pain points across different roles and tasks. Based on our findings, we also summarise key considerations and propose discussion highlights in typical use cases to guide future user studies and scenario-based HRI research in the USV domain.
It is worth noting that our study investigated both USV engineers' strategies for mentoring beginners and the learners' firsthand experiences, providing a comparative analysis of these perspectives. We hope that these insights will facilitate an open and critical exploration of potential use cases and scenarios in the context of HRI for USVs.

Ultimately, we first assess the usability of current USV systems from both engineer and beginner perspectives. Building on this, we investigate the mental models and adaptation strategies of beginners, as well as the approaches engineers use to mentor them. These insights shed light on how users develop operational understanding and what forms of support can enhance the learning process. From these findings, we then identify three representative use cases, each highlighting critical design considerations for HRI in USV operations. While these cases are not exhaustive—especially as technological advances may create new operational contexts—they capture common challenges and scenarios that are likely to persist over time.  These insights shed light on how users develop operational understanding and what forms of support can enhance the learning process.

The remainder of this paper is structured as follows. Section \ref{section2} outlines the research design, including participant recruitment, data collection, and data analysis procedures. Section \ref{section3} presents our empirical results, encompassing system usability assessment, beginners’ mental models and learning strategies, and three representative use cases that highlight critical operational challenges. The findings are discussed in Section \ref{section4}, leading to a broader rethinking in the form of design inspirations in Section \ref{section5}, which offers principles to guide the development of next-generation HRI design of USVs. Section \ref{limitation} acknowledges the study’s limitations, while Section \ref{conclusion} concludes by addressing our central research question and outlining directions for future HRI design of USVs. Supplementary materials, including interview protocols for engineers and end-users, are provided in the appendices.

\section{Research Design}
\label{section2}
The research design for this study is structured to explore the HRI challenges and use case scenarios in the context of USVs. 
Over the course of three months, we conducted in-depth interviews and observational studies with key stakeholders who had direct experience in at least one of the following areas: development, operation, or use of USVs.
This mixed-method approach allowed us to gather comprehensive insights from both beginners and experienced users. The study was approved by the university’s ethics committee, ensuring that all participant interactions adhered to ethical research standards. We describe our methodological approach in more detail below, including the participating roles, the data collection process, and the subsequent analysis used to derive meaningful insights regarding USV interaction design.

\subsection{Participants and Demographics}
The participants in this study were carefully selected to represent multiple stakeholders involved in the development, operation, and use of USVs. This diversity allowed us to explore the HRI challenges from different perspectives and gather varied insights into USV use cases. To recruit participants, the research team contacted a broad range of well-known companies and research institutions involved in USV development and operation via email. 
Snowball sampling\cite{naderifar2017snowball} supplemented purposive sampling to access niche stakeholder groups, with participant referrals continuing until theoretical saturation was achieved. All participants confirmed that there are no payments offered for their assistance following a verbal agreement between the interviewees and the researcher. After securing permission from the company/institution management, we were able to conduct semi-structured interviews with engineers and observe them as they completed scenario-based tasks using USVs. Through these institutions, we also engaged with current users of USVs to conduct further interviews, offering a comprehensive user perspective.
This approach facilitates a deeper exploration of the learning curves, mental models, and strategies employed in interacting with USVs. 

Finally, we recruited 9 stakeholders (all male), ranging from 22 to 48 years old, including 3 control engineers, 4 solution architects, and 2 end-users. All considered themselves to be professional operators, with an average of 4.83 years of hands-on experience (at least once per month for all participants, whereas experts once per week) operating USVs for professional missions. Our small sample size is not only due to the limited number of USV users compared to more common domains such as UAVs or AVs, but also a deliberate choice. By establishing continuous and productive relationships with our interviewees, we ensured a focus on the "depth" rather than the "breadth" of "thick data", as referred to Crouch et al. \cite{crouch2006logic} and Young et al. \cite{young2019examination}. 
In qualitative research, theoretical saturation—not sample size—governs analytical validity \cite{2013unsatisfactory, baker2018many}.
Previous studies on expert users have shown that a sample of 8 to 13 participants (e,g.: \cite{balarabe2021factors} and \cite{uzun2025suicide}) can yield acceptable results when saturation is achieved. Table \ref{tab:ethnographic list} shows the background of each participant involved in our research. In the remaining sections, we use pseudonyms with prefixes to refer to the role of participants: E-[pseudonym] refers to the Control Engineer, S-[pseudonym] refers to the Solution Architect, and U-[pseudonym] refers to the End-User. 


\begin{table*}[]
\centering
\caption{List of Participants Involved in this Ethnographic Study.}
\label{tab:ethnographic list}
\resizebox{\columnwidth}{!}{%
\begin{tabular}{llll}
\toprule
Pseudonym & Role                & Year of Experience & Experience Description \\ 
\midrule
E-Huang   & Control Engineer      & 5                      & Client task assistance; Design \& Integration; Quality Control \& Acceptance             \\
S-Zhu     & Solution Architect  & 5                       &    Supplier Management; Project Coordination; Algorithm provided           \\
E-Zhang   & Control Engineer  &      3.5                 &  Customer task assistance; Design \& Integration; Quality Control \& Acceptance          \\
E-Sheng & Control Engineer  &    2.5                    &   Customer task assistance; Design \& Integration; Quality Control \& Acceptance            \\
S-Zhong & Solution Architect  & 10  & Supplier Management; Project Coordination; Algorithm provided\\
S-Chen & Solution Architect & 6  &  Project Coordination; Algorithm provided; Customer task assistance\\
S-Sun & Solution Architect & 10 & Supplier Management; Project Coordination; Algorithm provided\\
U-Wu & End-User  & 0.5 & Undergraduate Student\\
U-Yang & End-User & 1  & Postgraduate Student \\
\toprule
\end{tabular}%
}
\end{table*}

\begin{table*}[]
\centering
\caption{Case list including time, personnel, and task names.}
\label{tab:case list}
\begin{tabular}{llll}
\toprule
Time        & Related Person & Task Name                     \\ 
\midrule
August 31st 2024 &  S-Zhu        & Harmful Algal Bloom Detection \\ 
February 27th  2025&       E-Huang    & Underwater Concealed Pipe Inspection\\ 
May 12th   2025&    E-Sheng            &   Post-Construction Hydrographic Survey \\ 
\toprule
\end{tabular}%

\end{table*}

\subsection{Data Collection}

In alignment with the immersive engagement approach recommended by Holloway and Galvin \cite{holloway2023qualitative}, we conducted a three-month immersive engagement at Company A from April to June 2024. During this period, we actively participated in daily operations, including observing field deployments, shadowing control engineers during mission planning and execution, attending design daily meetings with solution architects, and engaging in informal discussions with end-users during task execution. This deep involvement allowed us to gain first-hand experience and a contextual understanding of the operational environments, challenges, and workflows inherent to USV deployments. The insights and tacit knowledge gathered through these immersive activities—particularly the language/steps used by practitioners to describe their work—directly informed and shaped the development of our semi-structured interview protocol. This ensured the questions were grounded in the actual practices, terminology, and priorities encountered in the field, enhancing their relevance and applicability.

Subsequently, from June 2024 to May 2025, we conducted semi-structured interviews with engineers from multiple companies and institutions. The primary objective of these interviews was to understand the functionalities of existing systems, identify pain points, explore typical tasks, and gather guidance strategies for novice users. Each interview followed the same set of questions, ensuring consistency across all participants. Participants were encouraged to elaborate on their responses, and follow-up questions were posed based on their answers to gather more in-depth insights.

The median interview duration was 48 minutes (mean = 52.4 minutes), with most sessions occurring during working hours in the conference rooms of the participants’ companies. A detailed interview outline can be found in Appendix A. All interviews were recorded with the participants' acknowledge and consent, and measures were taken to ensure that transcripts were anonymised in line with ethical considerations.

To complement the interviews, we observed three typical use cases of USVs with the approve of the respective companies. This observational component aimed to supplement details that may have been overlooked during the interviews, particularly addressing issues related to common ground assumptions and memory biases. Although the operators were aware of the observation, this approach was considered appropriate given the exploratory nature of the research. Due to the sensitive nature of the data, no video recordings were made, and all observational content was recorded through handwritten paper-based notes.

We focused our field observations on three representative use cases corresponding to three different stakeholders. From a research design perspective, each use case represented a distinct category of USV task with standardised operational protocols, meaning that one observation per task type could provide meaningful insights into the broader category. From a practical standpoint, however, field observation involved exposure to sensitive data and client-specific workflows; several companies declined our request for site access for confidentiality reasons. Therefore, our observations were limited to the stakeholders who granted permission, and we prioritised scenarios with high relevant to common operational challenges discussed during the interviews.

Furthermore, detailed field notes were taken after each observation session, totalling 48 hours of observation time. Additionally, the researchers engaged with several current end-users for semi-structured interviews to understand their needs regarding USV interaction systems and their highlights of existing systems. These interviews were conducted online, and we went through with the interviewees their previous projects, identifying the main outcomes and findings of their projects, which lasted 15 to 30 minutes.
Their detailed interview outline can be found in Appendix B.
  
\subsection{Data Analysis}
The data consisted of field notes from informal interviews, transcriptions of formal interviews, and material provided by interviewees. Due to the entire study being completed offline in China, all data were collected in Chinese and then translated into English by the first author. Subsequently, the first author summarised the typical use cases by reviewing and annotating transcripts, using direct analysis \cite{parameswaran2020live} in qualitative research. As a secondary objective, the first author conducted inductive thematic analysis \cite{nowell2017thematic,creswell2016qualitative} to report the corresponding critical considerations of each use case and the mental models and learning strategies of USV beginners.
To enhance the trustworthiness of the qualitative analysis, the second author—an independent PhD student—was invited to cross-review the initial annotations. All coding disagreements were discussed and resolved through consensus between the two authors.

\section{Findings}
\label{section3}
\begin{figure}[h]
  \centering
  \includegraphics[width=\linewidth]{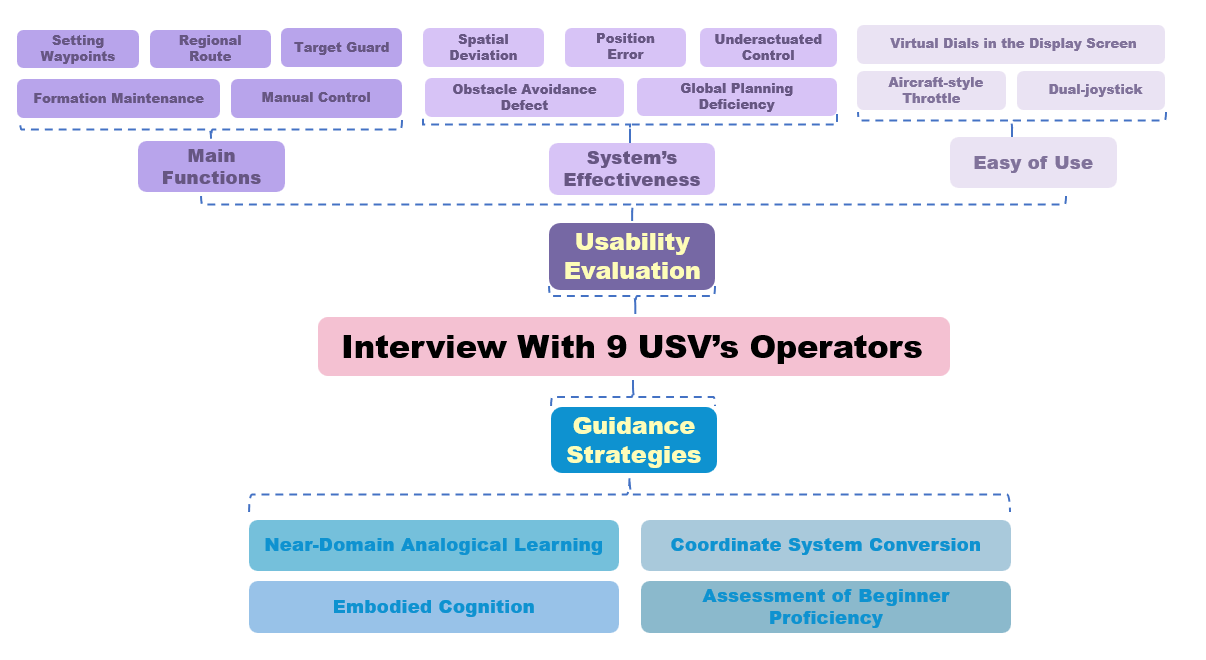}
  \caption{Thematic map illustrating key findings from the analysis.}
  \Description{Thematic map illustrating key findings from the analysis.}
  \label{Fig:themes map}
\end{figure}

This section presents our findings derived from interviews and observations conducted with stakeholders about USV. We begin by assessing the usability of the existing system, focusing on its effectiveness and user-friendliness based on expert and user feedback. Following this, we delve into techniques used by beginners to overcome initial difficulties, drawing insights from engineers' experiences to guide them. Finally,we summarise three typical use cases along with their considerations. Figure \ref{Fig:themes map} illustrates the relationships among the categories and themes identified during our analysis, and the finalised themes are the titles of the following subsections.

\subsection{Assessing Usability of Existing System}

The evaluation of the existing USV system focused on its usability, including the system’s effectiveness (in terms of functionality) and ease of use (in terms of interaction). Our assessment was based on in-depth interviews and field observations, which helped us gather insights from experts and users. Specifically, we explore what key functions are in experts’ perspective and how experts interact with key functions such as waypoint settings, automatic navigation, and manual control to understand the advantages and potential directions for improvement. This examination allowed us to identify both the strengths of the current system and potential directions for improvement. In doing so, we also uncovered several recurring usability challenges, including issues with interface design, control accuracy, and task reliability. Participants consistently highlighted pain points related to obstacle avoidance, underactuated control, and map accuracy. In this section, we delve into these findings, discussing the strengths and limitations of current USV systems and offering targeted recommendations for future enhancements.


\subsubsection{Main Functions}

Current USV systems employ a dual-mode operational framework combining automated navigation with human supervision. Like other unmanned vehicle sectors, automatic navigation involves setting waypoints, allowing the USV to navigate the predefined route while autonomously utilising obstacle avoidance algorithms. Control engineers frequently highlighted the critical importance of the waypoint function, particularly in navigating complex environments: "When we set waypoints, it allows us to plan an accurate route, which is especially important for avoiding obstacles or adjusting to real-time changes in conditions" (E-Wang). 

Most interviewees (8/9) mentioned that due to the specificity of current USV use scenarios (i.e. patrols, surveillance, mapping and rescue operations, etc.), the design of USV system is task-oriented, and many companies are required to customise USVs for specific tasks depending on environmental conditions. For instance, some systems allow users to manually define an area, after which the software generates a coverage route automatically. The summary of all the functions mentioned by the interviewees is as follows:

\begin{itemize}
    \item \textit{Setting Waypoints}: Widely considered as the most fundamental function, system generates routes based on operators manually setting waypoints, usually connected by a sequence of straight lines
    \item \textit{Regional Route}: After drawing a polygon area on the map interface or manually controlling the USV to trace a physical boundary, the system automatically generates routes that can cover that area, and the density of the routes (the spacing interval between adjacent paths) is adjustable to balance scanning resolution and operational efficiency
    \item \textit{Target Guard}: The system enables USV to maintain a specific distance from the target and protect it; usually used in aquatic environment safety and protection
    \item \textit{Formation Maintenance}:The system maintains a predefined formation by controlling each USV’s relative position to a designated leader, adapting speed and heading as needed. 
    \item \textit{Manual Control}: The system involves users directly operating the USV's speed and direction
\end{itemize}

During the interview process, four core functionalities—\textit{Setting Waypoints, Regional Route, Target Guard, Formation Maintenance}—were consistently classified under automated navigation paradigms, distinguishing them from manual control due to their reliance on the accuracy of electronic maps. These functionalities demonstrate the versatility of USV systems, but their effectiveness fundamentally depends on the accuracy of the electronic map. The precision of the electronic map is the foundation of waypoint programming, which, in turn, serves as the cornerstone of autonomous navigation.  Waypoint programming, as the cornerstone of autonomous navigation, requires sub-meter-level or even centimeter chart accuracy—a standard seldom maintained in dynamic aquatic environments.

\subsubsection{System’s Effectiveness}
\paragraph{\colorbox[RGB]{215,195,246}{Spatial Deviation}}
The interview data shows that the spatial deviation of electronic maps is mainly reflected in three aspects: hydrological seasonal changes (noted by 3 participants), nearshore tidal fluctuations (noted by 2 participants), and artificial temporary obstacles (noted by 2 participants). Inland river operators have generally reported significant impacts of the hydrological season (including flood season and dry season) on navigation baselines. As E-Huang stated, "The displacement of river boundaries during flood and dry seasons can reach 15-20 meters, but most electronic maps still use single-season data." Such deviations force operators to redefine coverage areas using the Regional Route function during seasonal transitions frequently. In urban inland river scenes, E-Sheng and E-Zhang emphasise that unforeseeable obstacles appear more frequently: "Coastal tides, fishing nets, and temporarily closed waterways are often not marked on maps and rely on visual observations along the riverbanks." 

To address unfamiliar river systems or areas where significant discrepancies exist between the map and the actual environment, many USV companies and institutions have incorporated a \textit{regional route} function into their systems. This feature allows operators to manually define a safe zone by controlling the USV, ensuring that the USV navigates within a more predictable and secure region. As E-Huang explained, this approach helps mitigate the risks of geospatial inaccuracies by allowing operators to establish boundaries before switching to autonomous control. When carrying out tasks in urban inland rivers, E-Sheng and E-Zhang both mentioned that they also have to use kayaks to follow USVs to avoid the aforementioned map error issues.


The selection of map tools also presents differences in environmental specificity. Companies engaged in inland operations generally use regular navigation maps, while companies engaged in offshore operations generally use Electronic Navigational Charts (ENCs)\cite{iho2000iho,international2018universal}-integrate hydrographic surveys and static hazard databases. To further elaborate on the limitations of standard mapping tools, S-Zhong noted, "For companies like ours, which operate in river and marine environments, you cannot rely on general-purpose mapping services like Gaode or Google Map. They are not designed for your lake or river systems; they focus primarily on road networks." As a solution, these companies often use ENCs, which provide more accurate static island and reef information, aiding in route planning for USVs. However, even with professional ENCs, there is still a problem of update lag in dynamic areas, such as tidal effects. As E-Zhang cautioned, "Each map version carries terrain bias, and we consider it as historical references, not absolute basis."  Moreover, S-Sun pointed out that nearshore operations must also accommodate temporary maritime directives issued by maritime bureaus, such as sudden channel closures or navigation restrictions. Consequently, even "autonomous" USVs require continuous oversight by remote operators (often one operator monitoring multiple USVs). Upon receipt of any such directive, the operator must immediately reconfigure mission paths or objectives to remain compliant.  

\paragraph{\colorbox[RGB]{215,195,246}{Position Error}}
Another recurring issue is Global Navigation Satellite System (GNSS) signal loss, which can occur in various situations, manifesting in three primary scenarios: remote areas with weak coverage, metallic bridge passages causing satellite signal loss, and electromagnetic interference in sedimentary coastal zones. Field data from S-Chen’s operations in the Yellow Sea’s saline-muddy regions revealed that even though complex terrains like islands and reefs are avoided, the unique landscape often causes electromagnetic interference or signal multipath reflections, leading to frequent positioning losses. Their current solution is to use operator's experience utilizing signal strength as an indicator to guide the USV, such as \textit{manually navigating} towards stronger signal areas, a method S-Chen described as "temporary and unreliable, introducing additional collision risks." To address error of position, inertial navigation systems (INS) integrating accelerometers and gyroscopes offer a technical solution by calculating real-time displacement and attitude during signal outages. However, cost barriers limit implementation—only 12\%(3/25) of surveyed USVs (primarily Company A’s premium models) currently deploy INS. In most cases, operators opt to wait for the ship to move to an area with stronger signals, as this is a simple and cost-effective approach (relying necessarily on the ship's collision protection due to cost constraints).

For mission-critical tasks such as nearshore surveying, S-Chen's team adopts a hybrid operational model where a manned mothership accompanies the USV to maintain mission continuity. This approach resolves three fundamental constraints inherent to autonomous systems. 
Firstly, centimetre-level navigation accuracy requires continuous calibration through shore-based reference stations, an infrastructure rarely available in dynamic coastal environments. By serving as a mobile reference station, the mothership maintains GNSS correction throughout the mission. Secondly, hydrographic surveys demand immediate course correction, as even small positional deviations can render a survey line unusable. As S-Chen explained, "Our bathymetric mapping demands less than 5 centimetres cross-track accuracy (from IHO (International Hydrographic Organisation) S-44 standard \cite{iho200844}). Automated systems take several seconds to replan, while mothership operators correct course deviations in real-time." Finally, the endurance limitations of the USV are mitigated by the mothership’s constant presence, as the USV’s battery life still falls short of fully replacing the mothership’s operational capacity. By having the mothership shadow the USV, the team ensures the mission can be completed without interruption, overcoming the endurance constraints of the USV. Cost-benefit analysis substantiates this approach. S-Chen quantified: "A 40-meter mothership incurs daily costs of ¥15,000–20,000. While USVs operate at 20\% efficiency, they reduce total expenses by 20\% through parallel operations." This pragmatic adaptation underscores the transitional phase of USV technologies—balancing automation potential with persistent infrastructural and economic constraints.

\paragraph{\colorbox[RGB]{215,195,246}{Underactuated Control}}
A frequently emphasised challenge is  'underactuation'—a fundamental characteristic that underlies many of the operational difficulties noted below.
The underactuated system refers to a scenario where the number of control inputs is fewer than the system's degrees of freedom. Unlike fully actuated systems (e.g., ground robots), USVs possess fewer control inputs than degrees of freedom (6-DOF vs. 3 controls).
Unlike vehicles where speed and steering inputs maintain relatively decoupled control relationships, USVs require coordinated multi-variable adjustments across propulsion, rudder angles, and ballast systems—all while compensating for six-degree-of-freedom environmental disturbances, including wave-induced heave/sway and wind-driven yaw moments. This makes USV control far more intricate and non-linear than terrestrial vehicle control. This underactuation affects the ability of USVs to perform actions such as hovering or maintaining a stable position.  Unlike terrestrial vehicles with direct throttle-brake linearity, USVs exhibit non-linear thrust-response characteristics requiring sequential gear transitions. As operational data from Company C demonstrates, abrupt course corrections from 30-knot cruising speeds demand expert throttle management: operators must first reduce propulsion to 40\%, execute rudder adjustments, then gradually reapply power—a process taking about 15 seconds versus less than 2 seconds compared with similar road vehicle manoeuvres. This mechanical inertia compounds positional control limitations. As S-Chen highlighted, “A USV cannot hover in place like an UAV. To ensure stability, operators must rely on their experience to stabilise the USV and make manual adjustments, especially when docking or dealing with rough conditions.” In Company C, skilled operators manually maintain positions within a 5-meter radius circle through thrust-vectoring techniques. Furthermore, unlike a car that can park steadily in place, docking a USV is more complex due to the vessel's sensitivity to environmental factors and underactuated nature. S-Zhong explained that the USV is constantly impacted by external forces during docking, making it difficult to position the USV steadily. The process requires the operator to find a method that minimises the interaction forces between the USV and the docking location, which utilises protective fenders to dissipate kinetic energy through controlled collisions. S-Zhong likened this challenge to positioning a small boat next to a large cruise ship, where the USV's movement is influenced by the surrounding conditions, such as wind or waves, necessitating careful control.

It can be understood that the docking task mainly depends on the operator's understanding and tolerance of collision risks, which depends on their experience, the complexity of the environment, and the static and dynamic characteristics of the USV itself \cite{CHENG2024109612,kim2020human,du2020review}. S-Chen further clarified that the current top algorithms in the industry still cannot meet the precise docking needs in complex environments. The reason is that, at present, USVs have a certain level of collision resistance, which reduces the actual impact of collisions. Therefore, the industry has not prioritised precise docking as a problem to be solved. This reflects the industry's pragmatic approach to solving short-term problems, but in the long run, with the increasing complexity of tasks and users' expectations for efficiency, this issue urgently needs to be addressed.

In the scenarios discussed above, manual control serves as a necessary backup mode to ensure the safety and efficiency of USV operations. For instance, in open ocean operations, S-Zhong’s team predominantly relies on planned missions for autonomous navigation. However, due to the underactuated nature of USVs, when approaching or leaving the port, especially when docking at the port, they typically resort to manual control via a remote controller, using their experience to steer the USV. Similarly, E-Zhang and E-Sheng mentioned that when map inaccuracies occur, they monitor the geographic environment while manually controlling the USVs. They also shared a strategy where manually circumnavigating suspect areas to calibrate geospatial offsets.

In certain cases, operators voluntarily choose manual control. E-Zhang explained that if precise coordinates are not critical and the task is confined to a visually observable area, manual control is preferred. Both E-Zhang and E-Sheng emphasised that the decision to use manual control often depends on the operator’s trust in the camera feed, which has a delay and limited field of view.
E-Zhang roughly estimated that the visually observable range in open environments was between 1000 and 1500 meters. In contrast, E-Sheng noted that although inland rivers are within this range, in larger bodies of water like Jinji Lake, the effective visual range expands to over 1500 meters due to reduced obstructions.
E-Huang mentioned that when the river width exceeds 200 meters, operators often struggle to distinguish the bow from the stern of the USV and thus prefer automatic control. For narrower rivers (less than 15 meters in width), E-Huang prefers manual control because it is more intuitive and faster than relying on a 2D map to plan waypoints. Specifically, throttle-rudder coordination is quicker and more responsive than algorithmic adjustments, while waypoint planning requires more experience to accurately judge the map points and scale conversion. S-Zhong agreed with E-Huang’s view that manual control is more sensitive, allowing for more detailed adjustments to the throttle and rudder.  Furthermore, manual control is sometimes required when the USV's onboard instruments need to be initialised or calibrated onshore. However, since manual control is subject to human limitations, it may not achieve the required task precision. For instance, when trying to navigate the USV along a straight line, visual errors prevent the USV from following an exact course. While automatic control may be less flexible, it typically achieves higher task completion accuracy.

U-Wu, a researcher who has used USVs for scientific data collection, shared his experience: "When sampling emergent algal blooms, preprogrammed paths prove useless. We manually 'hunt' hotspots through real-time water quality feedback." This operational pragmatism underscores manual control's enduring role as a scientific enabler in unstructured environments.


\paragraph{\colorbox[RGB]{215,195,246}{Obstacle Avoidance Defect}}
The operational demands of USV navigation exhibit marked divergence between inland and nearshore environments, necessitating distinct obstacle avoidance approaches. Companies A and B, predominantly serviced in Inland waterway systems (lakes/rivers), revealed three recurring obstacle scenarios: (1) entanglement risks from floating debris (e.g., fishing nets, aquatic vegetation, branches), (2) high-speed moving obstacles (humans or boats engaged in water sports), and (3) fixed obstacles (e.g., submerged rocks, bridge pilings). The inland environment is narrow, and there are many types of obstacles, emphasising precise and real-time dynamic obstacle avoidance. This places higher demands on the real-time path-planning capability of the algorithm. S-Zhu introduced a commonly used obstacle avoidance algorithm in the USV—Artificial Potential Field (APF): $F_{\text{total}} = F_{\text{attractive}} + F_{\text{attractive}}$. The algorithm generates attraction vectors $F_{\text{attractive}}$ toward target waypoints while applying repulsion forces $F_{\text{attractive}}$ proportional to obstacle proximity. The combination of these forces dictates the USV's navigation direction, enabling dynamic obstacle avoidance without preprogrammed path dependency. While the APF algorithm works well in ideal settings, especially in fixed obstacles, it is difficult to deal with high-speed moving targets, exposing its limitations. Target Unreachable\cite{rostami2019obstacle, you2022research}, local minima\cite{rostami2019obstacle, song2020path, luan2024path, sun2024path} and oscillation phenomenon\cite{9146273,sun2024path} remain major issues for the field of USV. Several interviewees identified three recurring failure modes necessitating manual takeover: oscillatory navigation patterns caused by inhomogeneous vegetation density ("The USV may circle endlessly through patchy kelp beds"); delayed response to high-speed obstacles due to perception range constraints ("Water skiers appear suddenly within 15-meter detection zones"); navigation paralysis in narrow channels where competing repulsive forces cancel propulsion vectors. 
These limitations necessitate manual intervention in most complex inland scenarios. Quantitative data reveals their prevalence in specific contexts: (1) floating debris forms permeable barriers (9 reported instances), (2) high-speed USVs breach 10-meter collision buffers (7 documented cases), or (3) channel widths contract below 1.5× USV beam dimensions (3 reported occurrences). S-Zhu further explains the root cause: "The system treats floating branches as impassable walls, though experienced operators know they can be traversed at reduced speed." 
This rigidity arises less from raw perception limitations than from the absence of adaptive engineering strategies for obstacle handling. While the current perception–planning pipeline detects and localises obstacles, it applies uniform, conservative avoidance rules that treat all objects as fully impassable. For instance, floating branches are categorised as solid walls, though experienced operators know they can be traversed at reduced speed. Without obstacle-type-dependent parameter adaptation in the APF algorithm, the system cannot exploit the operational flexibility available to skilled human pilots.

Contrastingly, nearshore operations within 22 nautical-mile zones leverage spatial flexibility to implement detour-priority strategies. As S-Chen explained, "Maritime collision regulations permit kilometre-scale diversions—we'd rather add 1.8 kilometres to a route than risk a 50-metre proximity." This operational philosophy capitalises on open-water manoeuvrability while addressing dynamic challenges like tidal currents. Moreover, nearshore operations involve more complex conditions due to tidal fluctuations exceeding 2-meter vertical variations, dense maritime traffic patterns (including commercial ships and fishing ships), and strict compliance with International Maritime Organisation (IMO) regulations. In this context, thanks to the Automatic Identification System (AIS) transponders with Very High Frequency (VHF), which facilitate communication between vessels in nearshore environments, operators can share navigational information to avoid collisions. S-Chen explained the potential advantage of AIS: "AIS allows vessels to communicate their locations and courses, helping USVs navigate through congested waters and avoid collisions. In this case, human intervention is not always necessary, as communication between vessels takes precedence over algorithm-based avoidance." A critical capability was validated during a 2023 incident documented by S-Chen:  When a cargo vessel operator fell asleep at the helm near the Yangtze River estuary, violating standard collision-avoidance assumptions, the operator of the USV used VHF to wake up the sleeping captain, thereby avoiding this collision situation. This case underscores the superiority of AIS-enabled direct communication over predictive algorithms alone.

\begin{figure}[h]
  \centering
  \includegraphics[width=0.6\linewidth]{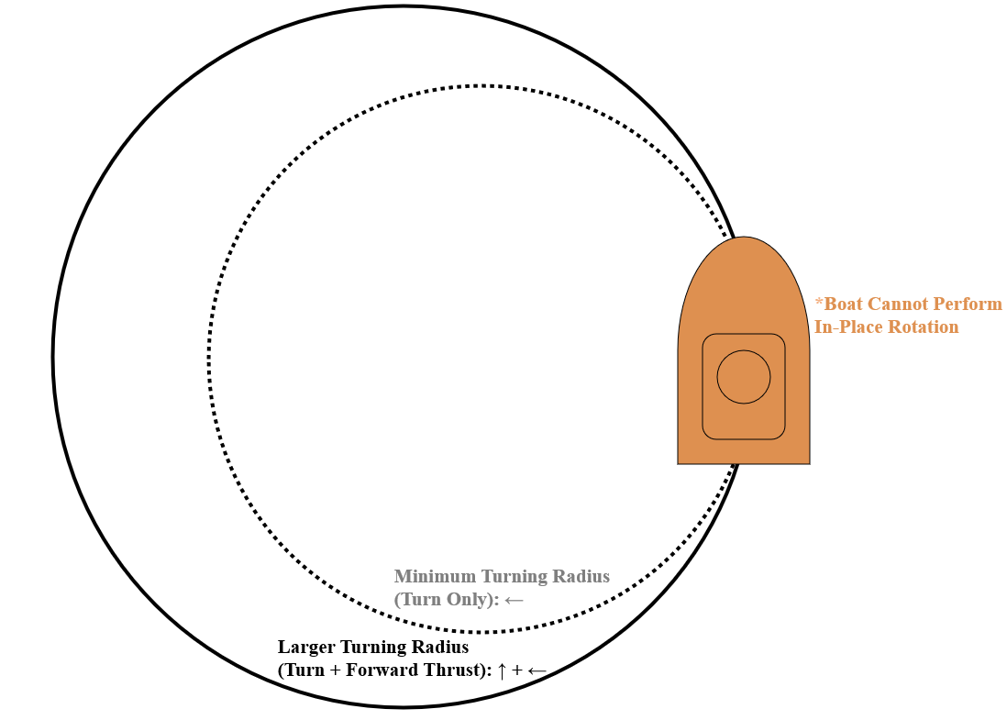}
\caption{The turning radius of USVs.}
  \Description{This Figure illustrates the turning radius of USVs.}
   \label{Fig:turning}
\end{figure}

\paragraph{\colorbox[RGB]{215,195,246}{Global Planning Deficiency}}
All USV systems employ a global path-planning strategy, which involves multiple factors and differs significantly from simpler path-following algorithms, such as those used in robotic vacuum cleaners. S-Chen emphasised that, unlike vacuum cleaners that prioritise covering the maximum area with the shortest path, USVs must consider obstacle avoidance as the primary concern, followed by adherence to their own dynamic models, such as the required turning radius. In contrast to robotic vacuum cleaners, USVs cannot pivot steering, especially in environments with water currents and other dynamic interferences.  Diagram \ref{Fig:turning} illustrates the turning constraints of a USV: it cannot rotate in place; In an idea environment, when a left turn command is given without forward thrust, the USV has the minimum turning radius (dashed line); When both a left turn and forward thrust are applied, the turning radius increases (solid line). The turning radius $R = \frac{V}{\omega}$, where V refer to sailing speed and $\omega$ refers to rotational angular velocity of the USV.

Accordingly, \textit{Regional Route} is not solely based on the shortest route; instead, it also incorporates energy efficiency. For example, depending on the USV's speed (e.g., 12 knots, 10 knots), the allowable deviation from the planned path before corrective action is needed changes. S-Chen outlined that deviations are permitted up to a set distance—such as 20 meters or 15 meters(taking a speed of 12 or 10 knots as an example)—depending on the speed, as long as the USV remains within the optimal energy consumption range. Therefore, the main objective is not simply to minimise travel distance but to achieve the lowest energy consumption possible.
To address energy concerns, the system avoids inefficient behaviours like decelerating and then accelerating, which can significantly drain power. Instead, the system chooses a slightly longer route or increases the turning radius to maintain an energy-efficient path. Even if deviations occur during straight-line navigation, the USV will return to the original planned route only after ensuring that the turning radius and other constraints are met, thereby balancing navigation accuracy and energy consumption.

To illustrate the range of functionalities available across different USV systems in our interview, we present Table \ref{tab:functions list}, which maps each interviewee to their respective company/institution and the specific functions their USVs are designed to perform.

\begin{table}[]
\centering
\caption{List of Companies and their Functions in our Interviews}
\label{tab:functions list}
\resizebox{\columnwidth}{!}{%
\begin{tabular}{|l|l|l|l|}
\toprule
\textbf{Company/institutions ID} & \textbf{Their Function}                                                                              & \textbf{Related Interviewee} &\textbf{
Operational Domain} \\ 
\midrule
A                     & Setting Waypoints, Manual Control, Regional Route                                                    & E-Huang, E-Zhang, E-Sheng    & 
Inland Waterways \\ \hline
B                     & Setting Waypoints, Manual Control                                                                    & S-Zhu                    &    
Inland Waterways\\ \hline
C                     & Setting Waypoints, Manual Control, Target Guard, Formation Maintenance, Regional Route & S-Zhong           &      Nearshore Zones     \\ \hline
D                     & Setting Waypoints, Manual Control, Target Guard, Regional Route                                         & S-Chen            &     Nearshore Zones     \\ \hline
E   & Setting Waypoints, Manual Control, Target Guard, Formation Maintenance, Regional Route & S-Sun           &      Nearshore Zones     \\ \hline
\toprule
\end{tabular}%
}
\end{table}

\subsubsection{Ease of Use}
The systemic constraints outlined above highlight the significant dependence on human oversight, underscoring the need for an intuitive and effective manual control system. However, manual control introduces its challenges, particularly in terms of human-machine interaction. This subsection examines the user experience during the operation of USVs, with a specific focus on manual control tools. The ease of use of these tools—whether joysticks, mice, keyboards, or custom controllers—directly impacts the operator's ability to control the vehicle effectively, comfortably, and efficiently. 

\begin{figure}[h]
  \centering
  \includegraphics[width=\linewidth]{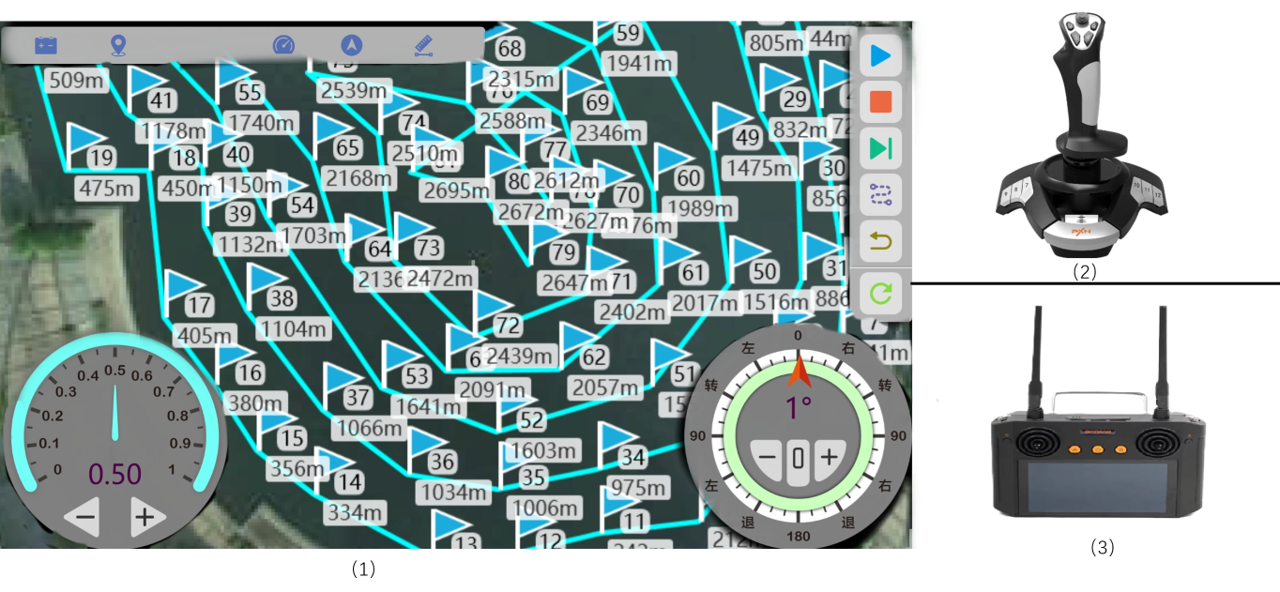}
  \caption{The Three Manual Interface Types Include: (1) The Interface Provided by a Certain Company; (2) Aircraft-Style Throttle Control; and (3) Dual-Joystick Gamepad.}
  \Description{Three Interface Types: (1) The Interface Provided by a Certain Company; (2) Aircraft-Style Throttle Control; and (3) Dual-Joystick Gamepad.}
   \label{Fig:interface}
\end{figure}

Contemporary USV control systems employ tri-modal manual interfaces: (1) virtual dials in the display screen (0-1 speed scale, 0\textdegree-360\textdegree directional control), (2) aircraft-style throttle control, and (3) dual-joystick gamepad. Figure \ref{Fig:interface} presents a screenshot of the software interface and the controller, illustrating these interaction methods.

Field observations reveal distinct operator preferences shaped by environmental and ergonomic factors. For instance, compared to the tablet device, some surveyed operators(2/9) prefer laptop interfaces due to their long battery life(more than 5 hours runtime), larger display screen and rain-resistant mouse inputs. However, the lack of portability in laptops is considered a significant drawback. In contrast, almost all surveyed operators (7/9) express dissatisfaction with the virtual dials, particularly because they do not feature a reset mechanism. The absence of this feature makes the system less intuitive, requiring operators to manually adjust settings, which can be cumbersome, especially in fast-paced situations. Additionally, due to latency in the camera transmission, operators must often predict and anticipate the movements of the USV in advance. As a result, controllers equipped with a reset mechanism—allowing for quicker adjustments and more intuitive control—are preferred by operators. It is also worth noting that, compared to virtual dials on display screens, which are not limited by communication distance when using cloud-based systems, physical controllers based on short-range signals have a range limit (typically 1-2 kilometres, depending on the brand and environmental conditions). This range limitation means that controllers are generally used within the operator’s line of sight (hereinafter referred to as the "shore-based perspective").

Using a controller from the shore-based perspective has several advantages. First, in emergency situations—such as when a USV fails to dock—the operator can immediately take corrective actions without waiting for the transmission and processing of electronic signals. Second, controllers tend to have lower latency, which is crucial for high-precision tasks such as docking or unmooring operations. In these scenarios, the ability to respond quickly can significantly improve the accuracy and safety of operations. Additionally, the shore-based perspective allows operators to better respond to dynamic environmental changes. In complex waterways, operators can observe real-time shifts in water currents, surrounding obstacles, and other environmental factors, which enables them to make swift adjustments. This sensitivity to environmental changes is one of the major benefits of shore-based operations.

Furthermore, the shore-based perspective takes full advantage of human visual and auditory strengths. Operators can not only clearly observe the USV’s status but also detect sounds from the surrounding environment, such as nearby ships’ horns or changes in water flow. In this paradigm, it utilises multimodal human perception, where operators integrate visual cues (such as wave patterns and obstacle shadows) with auditory signals (including engine sound and water turbulence), which onboard sensors typically filter as noise, providing a more comprehensive understanding of the environment and enabling better decision-making. As S-Zhong concluded, "Our eyes and ears perceive what algorithms ignore—the ripple warning of submerged logs, the wake patterns of hidden currents. No camera yet replicates that [wisdom]."

Despite these advantages, the shore-based perspective has limitations. As the distance between the operator and the USV increases, or when the view is obstructed, the ability to observe fine details diminishes. This is particularly true when judging the position of the USV, environmental changes, or underwater obstacles at long distances. Beyond the operator's eyesight, environmental lighting conditions also play a significant role. For example, bright sunlight or low-light conditions at night can hinder the operator's ability to clearly observe the USV and its surroundings, especially when the water's surface obstructs visibility. These challenges make the shore-based perspective more limited in certain conditions. To mitigate these limitations, cameras equipped with thermal imaging or night vision technologies can enhance visibility by providing high-resolution images and multi-angle views.

\begin{figure}[h]
  \centering
  \includegraphics[width=0.8\linewidth]{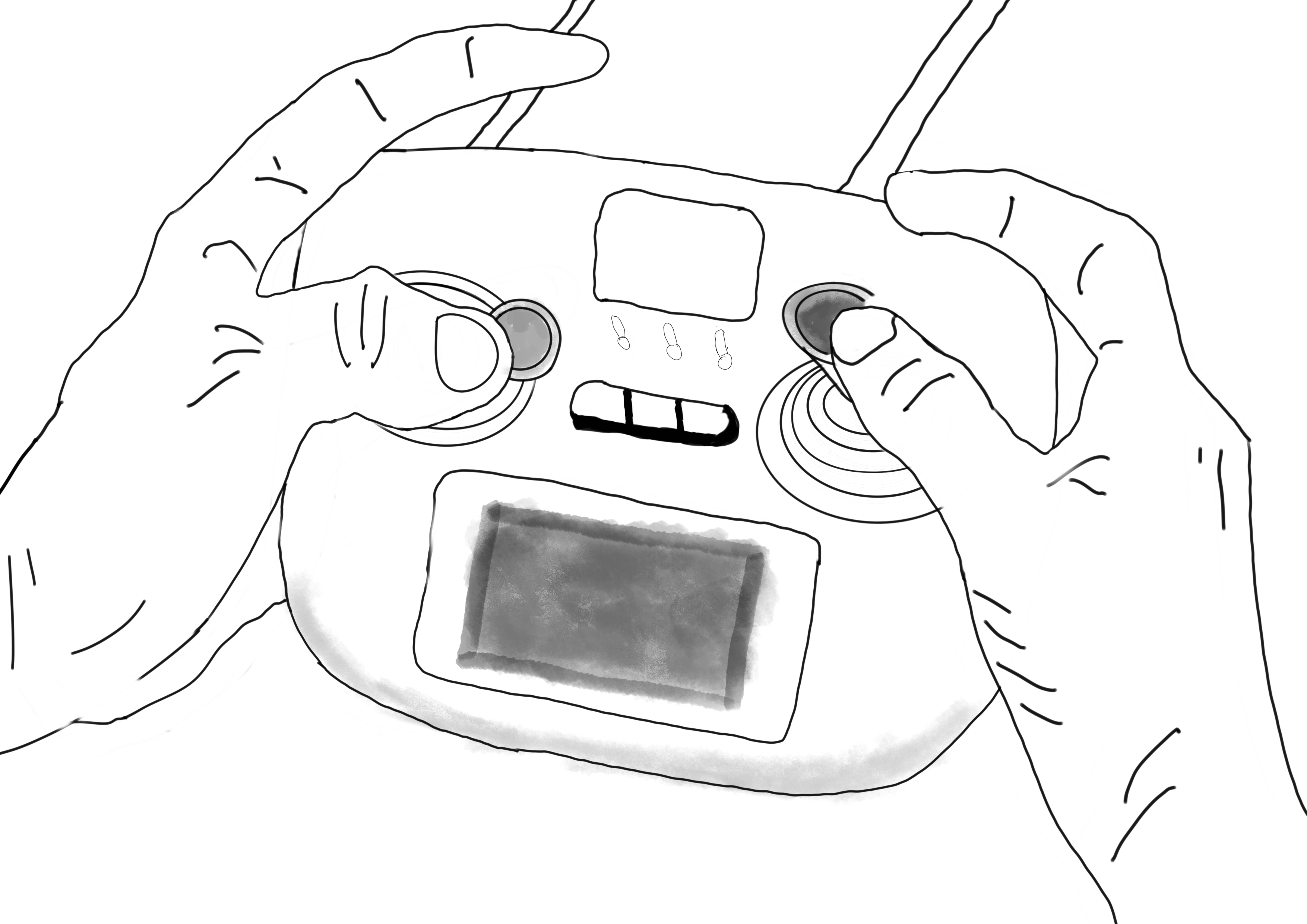}
  \caption{Dual Joystick Configuration: The Left Joystick Controls Left–Right Direction, and the Right Joystick Controls Forward–Backward Throttle.}
  \Description{the dual joystick}
   \label{Fig:controller2}
\end{figure}

\begin{figure}[h]
  \centering
  \includegraphics[width=0.5\linewidth]{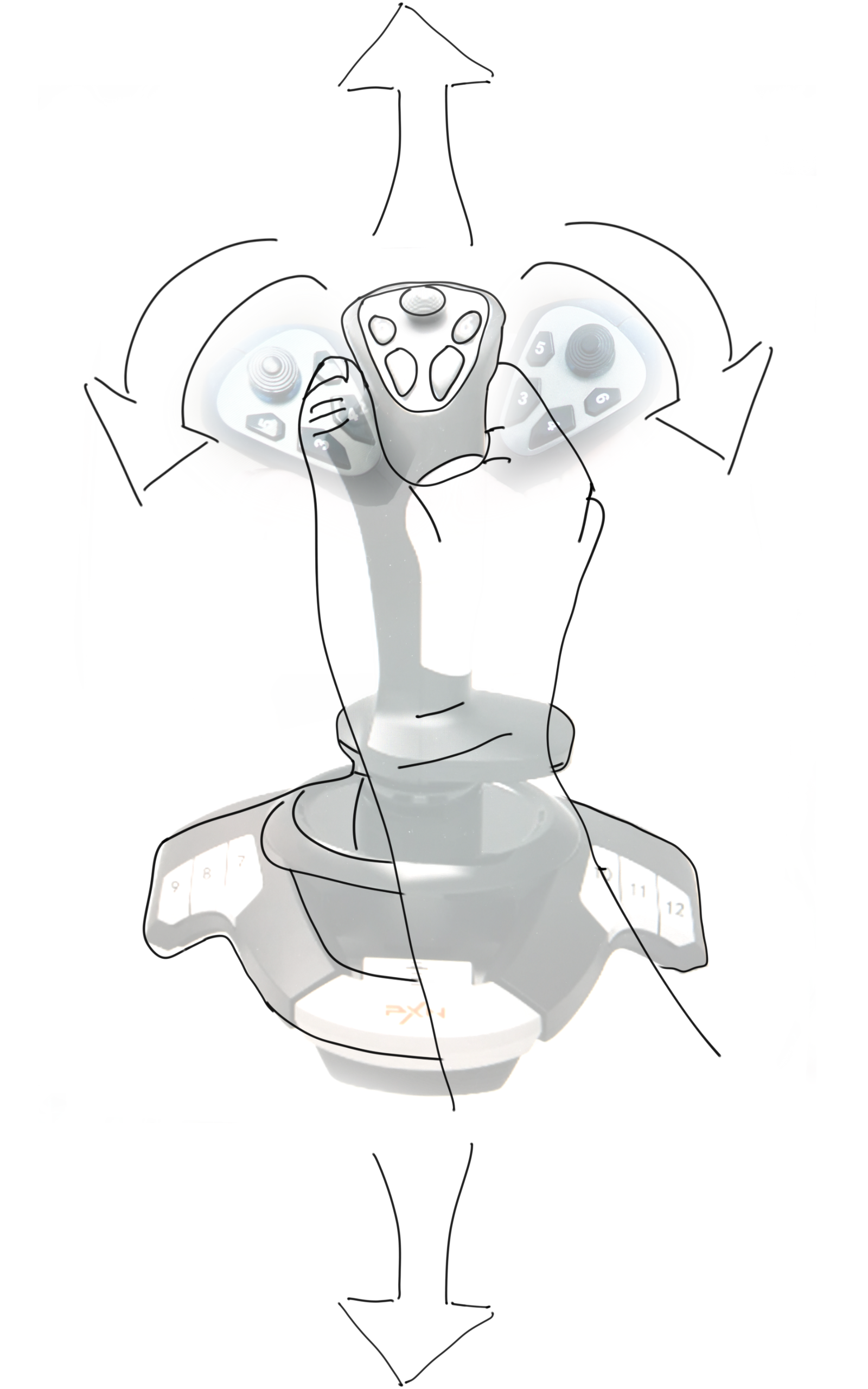}
  \caption{Aircraft-Style Throttle Control: Operators Control Speed by Pushing the Stick Forward or Backward, and Control Direction by Rotating the Stick Left or Right.}
  \Description{aircraft-style throttle control}
   \label{Fig:controller1}
\end{figure}

Current controllers used for USV operation generally come in two types: aircraft-style throttle quadrants and dual-joystick gamepad configurations. The first versions of USVs from all leading manufacturers (A, B, C, D, E) featured the dual-joystick gamepad configuration. E-Zhang pointed out that, unlike UAVS, where the left joystick controls rotational yaw (±180 $^{\circ}$) and vertical thrust (Z-axis) while the right joystick manages forward-backward (X-axis) and lateral movement (Y-axis), USVs are designed with twin propulsion systems. Consequently, in USV, one joystick deflection for azimuth control (left-right) and one joystick displacement for thrust modulation (forward-backward throttle), see in the Figure \ref{Fig:controller2}. However, while dual-joystick configurations widely used in many fields, they pose a steep learning curve for beginners unfamiliar with gamepads. S-Chen further explained that users found it confusing and had difficulty distinguishing between the joystick responsible for speed control and the one for directional control. Since many users have given unfriendly complaints on this controller, prompting a shift to a more traditional single-stick controller by manufacturers (D, E) that resembles the controls of vehicles or aircraft. This revised design provides tactile feedback and enables the operator to control speed by pushing the stick forward or backward, while rotational movement of the stick adjusts the direction, see in the Figure \ref{Fig:controller1}. To minimise unintended steering during forward or backward movements, the steering function was exclusively assigned to the stick's rotational action.

This shift to a single-stick throttle control is favoured for its intuitive and ergonomic design. The majority of USV operators, however, have experience driving vehicles, making the single-stick design more familiar and comfortable. As U-Yang, a user with a year of USV experience, noted, “Dual joysticks give me a sense of directly controlling the vessel, but I need some time to adapt to the sensitivity of the controls.” E-Huang highlighted the benefit of joysticks in complex environments: “In complicated situations, I feel like the joystick provides more immediate feedback. But in a lab or indoor setting, mouse can allow quicker adjustments.” 

Interviewees provided valuable feedback for a more intuitive interface and interactive system to reduce cognitive load. As S-Zhu pointed out, "When I have to manually adjust the USV's direction and speed, I often have to focus entirely on the control interface, leaving little [mental bandwidth] for monitoring the environment." S-Chen added that although some USVs are equipped with side cameras for multi-angle perception, their effectiveness is limited by users' attention span and lack of professional expertise, leading to inefficient camera use. Experienced captains may instinctively know when to look right or left, but less experienced users often struggle to manage multiple perspectives simultaneously. Building on this, S-Zhu suggested that future systems could incorporate a hands-free interaction method, as operators in outdoor environments are often occupied with other tasks.

\subsection{Mental Models and Learning Strategies of Beginners}

New users of USVs, called beginners in this section, despite having foundational knowledge of aquatic environments, often struggle with controlling the USV, especially when it comes to making precise adjustments or responding to unexpected technical issues. According to the engineers we interviewed, although these novice users are generally familiar with the tasks, their lack of experience with USV controls leads to slower response times and difficulties in dealing with technical aspects, such as adjusting speed, direction, or dealing with sudden technical malfunctions. In addition to the technical challenges, USV operators must also anticipate and react to environmental factors that could affect the USV's operation. For example, sudden changes in river conditions, such as turbulence, whirlpools, or floating debris, can interfere with the USV's path and performance. Operators are trained to identify these hazards in advance and plan routes that avoid or bypass them. Wind, especially in open river sections, may also alter the USV's direction, and operators must adjust the heading accordingly. Similarly, debris like fishing nets or water plants, or even floating obstacles in the waterway, can obstruct the USV’s path or entangle the propeller. In such cases, operators need to steer clear of these obstacles. In emergency situations, such as low battery levels or technical malfunctions, operators rely on their understanding of the river's flow and terrain to select the shortest or safest return route. S-Zhong highlighted that many enterprise-level users (such as governmental agencies and corporations, as mentioned earlier) are focused more on the outcome of the mission rather than the real-time state of the USV. This goal-oriented, mentality affects their approach to USV operations. For instance, in mapping tasks, professionals quickly assess the quality of collected data, while in patrol security tasks, their main concern is whether the system can effectively track and identify targets-particularly in emergencies where unknown USVs may enter the security area. These insights reveal the differences in mental models between engineers and new users, with the latter relying more heavily on simplified or trial-and-error strategies to navigate complex tasks (S-Chen). Understanding these challenges is essential for improving the USV learning experience. This section reflects on these difficulties and the strategies employed by both engineers and beginners to guide the learning process and achieve successful task completion.


\subsubsection{Near-Domain Analogical Learning}

A critical strategy employed by engineers to support beginners is analogical learning, which involves drawing parallels between USV operation and other domains that users are already familiar with. This approach taps into the concept of cross-domain learning, where skills and knowledge from one domain (e.g., UAV operation) are transferred to another (e.g., USV operation). By transferring their UAV operation experience to USV operation, users can rely on analogous control logic, which reduces cognitive load and accelerates learning. For example, users familiar with controlling a UAV's altitude, speed, and direction can draw on similar mental models when adjusting the speed and direction of a USV. As E-Huang's empirical observation-

\begin{quote}
``When I handed over the USV to a beginner, I told him it's similar to a UAV. They may already have a concept in mind: UAVs can take photos, plan routes, fly automatically, and be controlled with a joystick controller. This gives them the confidence to operate a USV.'' (E-Huang)
\end{quote}

The observations of new users are in line with engineers' statements:

\begin{quote}
``I think this system is fine because, from my perspective, it doesn't really affect the operation much. You just need to think about it briefly, like when playing a racing game, and you can control it easily. If it were designed with an absolute direction instead of a relative one, you know, it might actually make the operation harder.'' (U-Wu)
\end{quote}

\begin{quote}
``I think it's relatively easy to get started with, similar to a toy car. Sometimes I feel like playing Mario Kart.'' (U-Yang)

\end{quote}

The principle of analogy, rooted in educational psychology, suggests that when two domains share structural or operational similarities, learners can apply experiences from the familiar domain to more easily understand and master the new domain \cite{Lobato2006AlternativePO}. In the case of UAVs and USVs, both systems share key features, such as spatial navigation, control interfaces, and system responsiveness. These similarities make the analogy between the two domains more effective. In the case of cars and USVs, both involve directional control and speed adjustments, though the specific mechanics differ, making the initial learning curve less steep for individuals familiar with one type of vehicle.

By applying analogy principles, beginners can bridge the gap between their existing knowledge ( UAV and car control) and the new domain of USV operation, enabling them to learn more efficiently and with greater confidence. As a result, analogical learning not only accelerates the adaptation process but also fosters a deeper understanding of how control mechanisms across different vehicles function, further improving the usability and accessibility of USVs for new operators.

\subsubsection{Coordinate System Conversion}
Due to map inaccuracies, users often need to manually establish a safety zone in advance and develop proficiency in scale conversion. One major challenge is the reliance on GPS-based point marking, which requires operators to reverse-calculate positions—a process that demands experience and spatial awareness. This coordinate system conversion requires operators to mentally reconcile three interdependent spatial frameworks: 1) the 2D map interface's Cartesian grid, 2) the USV's egocentric movement vectors, and 3) the environmental topology (shorelines, currents, and obstacles). Enhancing spatial perception is therefore critical, as it directly influences the cognitive processes involved in spatial orientation and navigation, particularly in complex environments. As E-Huang's empirical observation—

\begin{quote}
"Manually plotting waypoints requires constant scale conversion—for example, a 2 cm measurement on the map might correspond to 20 meters on the water. [However,] only experience can truly guide this process."
\end{quote}

To assess whether beginners have mastered map reading and scale conversion, E-Huang suggested a practical evaluation method: asking novice operators to plot a straight-line route as close to the shore as possible while maintaining a safe boundary (typically considered to be two meters from the shore). If the planned route is safe, it indicates that the operator has acquired the necessary skills for route planning and USV control.

It epitomises the cognitive burden of proportional reasoning under uncertainty. Research from Verde et al. \cite{verde2018spatial} also indicates that reading maps effectively requires the ability to interpret small-scale representations and translate them into real-world, large-scale spatial positioning which varies from environmental complexity and individual spatial abilities. When the "up" direction on a map aligns with the "forward" direction in the environment—matching the perspective in which the spatial layout was learned—users make faster and more accurate judgments. This alignment effect highlights the importance of designing interfaces that support intuitive spatial reasoning.

\subsubsection{Embodied Cognition}
\label{subsubsection:Embodied cognition}

The theory of embodied cognition posits that cognitive processes are deeply rooted in the body's interactions with its physical environment—a paradigm fundamentally disrupted in remote USV operations\cite{wilson2002six}. Remote operators face a sensorimotor decoupling challenge: their spatial cognition, motor coordination, and situational awareness must operate without the proprioceptive, tactile, or auditory cues inherent to onboard navigation\cite{nostadt2020embodiment}. The observed difficulty novice operators face when controlling USVs with the aspect of astern face operator (relative to their viewpoint) reveals fundamental conflicts between embodied spatial reasoning and interface-mediated control paradigms. E-Sheng's empirical observation—

\begin{quote}
"Beginners instinctively expect forward joystick movements to correspond with the USV's bow direction, creating cognitive dissonance when orientations are reversed." (E-Sheng)
\end{quote}

It epitomises the sensorimotor mismatch inherent in remote marine operations. When the USV's bow faces the operator, control inputs require mental coordinate transformation (forward = backward in vehicle motion). This violates the natural proprioceptive mapping humans develop through embodied interactions with vehicles (cars, bicycles), where control directionality aligns with observed motion, forcing operators to maintain dual mental models: interface logic vs. environmental reality. Inconsistent control direction disrupts the egocentric spatial framework by requiring operators to mentally rotate obstacle positions (a 180° coordinate transformation), continuously shift perspectives between operator-centric and USV-centric references.

\subsubsection{Assessment of Beginner Proficiency}
Determining whether beginners have internalised USV control protocols requires evaluating both procedural compliance and adaptive problem-solving capabilities. S-Chen's cautionary observation—

\begin{quote}
    "Although the system provides a waypoint-setting function on the map, we do not allow beginners to set waypoints in the first attempt. Beginners often make errors, such as accidentally placing extra waypoints, causing the USV to deviate from the intended route. When they notice the discrepancy—realising that the travel time does not match expectations—they may attempt to modify the route without stopping the vessel first. Even if instructed to pause navigation and exit the auto mode before making adjustments, they may not strictly follow these guidelines. Since they perceive no immediate consequences—believing that mistakes carry no real cost—they are more likely to experiment freely rather than adhere to a structured procedure. This lack of strict operational discipline is one of the biggest challenges in training new users."
\end{quote}

Standardised training procedures emphasise stopping the USV before modifying waypoints to ensure system stability and navigation accuracy. The hardware limitations of unmanned surface vehicles also impose constraints. Even if the software can process new waypoints instantly, the mechanical response of the steering and propulsion systems takes time. For instance, the rudder requires a certain period to complete a turn, and the propulsion system needs time to adjust its power output. If a waypoint is modified while the vessel is travelling at high speed, these response delays may cause the USV to deviate from its intended course or even lead to hazardous situations. Furthermore, the maritime industry follows strict safety protocols based on extensive practical experience and risk assessments, reinforcing the necessity of proper procedural adherence.

However, effective training goes beyond procedural instruction. As E-Zhang and E-Sheng emphasise:

\begin{quote}
    "Manuals explain 'how,' but only practice teaches `when'."
\end{quote}

This highlights the gap between theoretical knowledge and practical expertise. Their company’s training protocol mandates supervised hands-on training, requiring operators to complete both manual control and autonomous missions before being allowed to operate USVs independently. While written manuals provide foundational guidelines, real-world navigation requires intuitive decision-making that can only be developed through experience.

Given these challenges, E-Zhang and E-Sheng acknowledge the potential of augmented reality (AR) as a training aid. Their moderate endorsement (3/5 on a Likert scale) reflects cautious optimism about AR’s ability to bridge the knowledge gap by dynamically visualising adjustments and providing real-time situational feedback during operation. By integrating AR-based guidance, training programs could reinforce procedural adherence while helping operators develop the situational awareness necessary for effective USV control.



\subsection{Typical Use Cases and Highlights}
\begin{figure}[h]
  \centering
 \includegraphics[width=\linewidth]{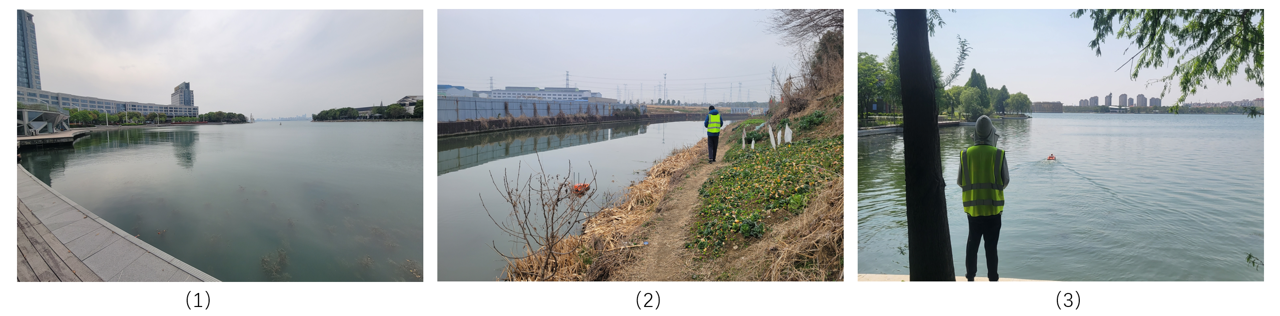}
  \caption{Three Typical Use Cases: (1) Harmful Algal Bloom Detection; (2) Underwater Concealed Pipe Inspection; (3) Water Quality Monitoring.} 
  \Description{Three Typical Use Cases: (1) Harmful Algal Bloom Detection; (2) Underwater Concealed Pipe Inspection; (3) Water Quality Monitoring.}
  \label{Fig:usecase}
\end{figure}

This section examines three operationally significant USV application scenarios derived from interviews with engineers, selected for their HRI complexity and environmental representativeness. Use cases, as noted by Dzida \cite{Dzida}, are valuable for defining context (from the user’s perspective), functionality (from the designer’s perspective), and for identifying problem domains during interaction design. Each case was validated through a six-hour-long field observations complementing semi-structured interviews, enabling triangulation of semi-structured interviews with empirical behavioural patterns. The selected scenarios—harmful algal bloom detection, submerged concealed pipeline inspection, and water quality monitoring—epitomize distinct navigation challenges: spiral area coverage optimization, shore-parallel manual trajectory control, and midline route adherence, respectively, which revealing three core tension points: automated efficiency versus manual precision, sensor limitations versus environmental variability, and interface complexity versus operational urgency. Field documentation includes georeferenced site diagrams (Figure \ref{Fig:usecase}), which show the environmental parameters of the site: waterway traffic, terrain, aquatic plant density, etc.

\subsubsection{Scenario One: Harmful Algal Bloom Detection}
\begin{figure}[h]
  \centering
 \includegraphics[width=\linewidth]{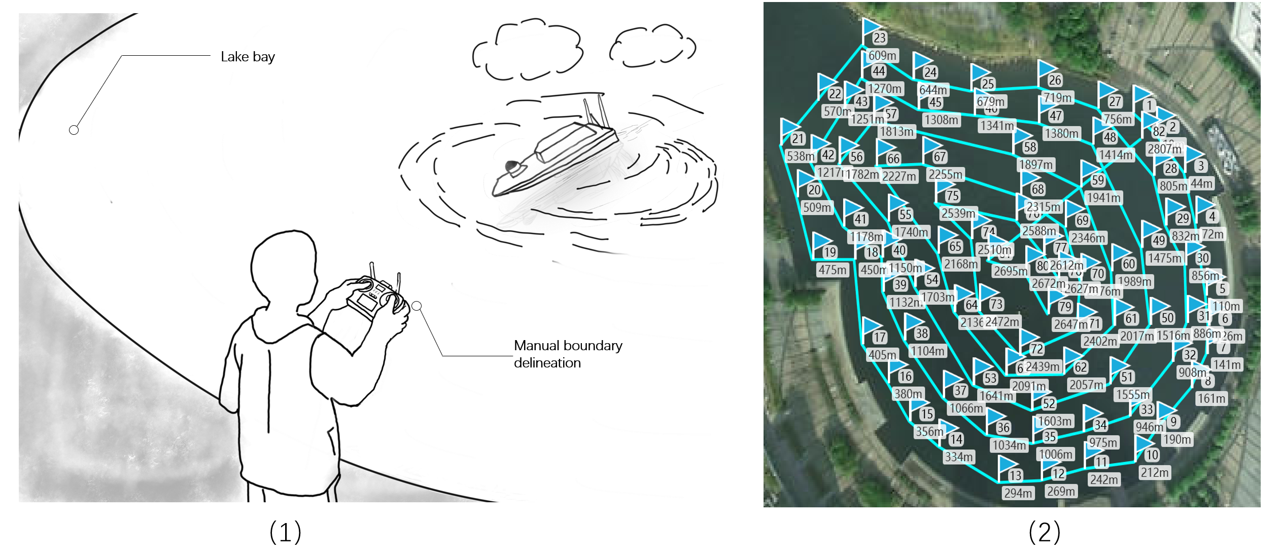}
  \caption{Operational scenario and task schematic of the harmful algal bloom detection: (1) Manual boundary delineation, where the operator pilots the USV along perceived shorelines to compensate for outdated or unreliable hydrological maps. (2) A spiral waypoint pattern starting from the shoreline.} 
  \Description{It sketches the scenario one: Harmful Algal Bloom Detection. (1) Manual boundary delineation, where the operator pilots the USV along perceived shorelines to compensate for outdated or unreliable hydrological maps. (2) A spiral waypoint pattern starting from the shoreline.}
  \label{Fig:S1}
\end{figure}

\paragraph{Scenario Description} This mission focuses on detecting and mapping harmful algal blooms in freshwater systems, typically in rivers or lakes prone to seasonal eutrophication. Operators deploy USVs equipped with multispectral cameras, chlorophyll fluorescence sensors, and water quality probes (pH, dissolved oxygen, and turbidity). The target area often spans dynamically changing water bodies where historical maps become unreliable during flood seasons due to shifted shorelines and submerged vegetation. The operation begins with manual boundary delineation: operators pilot the USV along the perceived edges of the waterbody to define the detection zone, compensating for outdated or inaccurate hydrological maps (see Figure \ref{Fig:S1} (1)). This is followed by a spiral waypoint pattern starting from the shoreline and expanding outward to ensure full coverage, particularly critical in irregular bays or zones with floating algal mats. Once the spiral path is validated (see Figure \ref{Fig:S1} (2)), the USV switches to autonomous mode. Environmental challenges include dense surface scums that obstruct propulsion, biofilm interference with optical sensors, and sudden wind-driven algal aggregations. 

\paragraph{Discussion Highlights} The manual boundary survey phase addresses a critical gap in automated systems: flood-induced hydrological changes that render preloaded maps obsolete. Operators leverage visual cues (waterline stains on vegetation, sediment plumes) and tactile feedback from the USV’s collision alerts to refine detection zones. For instance, abrupt deceleration patterns may indicate submerged obstacles not visible on surface scans. However, this introduces cognitive friction—operators must mentally reconcile conflicting data streams (historical GIS layers vs. real-time camera) while avoiding overcorrection from transient anomalies like floating debris. Compared to pure manual plotting, the spiral waypoint generation algorithm may provide more efficient interaction capabilities: when operators place 3 to 5 shoreline anchor points, the system calculates optimal logarithmic spiral parameters through iterative curve fitting, simultaneously adjusts the spiral spacing automatically based on the concentration of blue-green algae.

\subsubsection{Scenario Two: Underwater Concealed Pipe Inspection}

\begin{figure}[h]
  \centering
 \includegraphics[width=\linewidth]{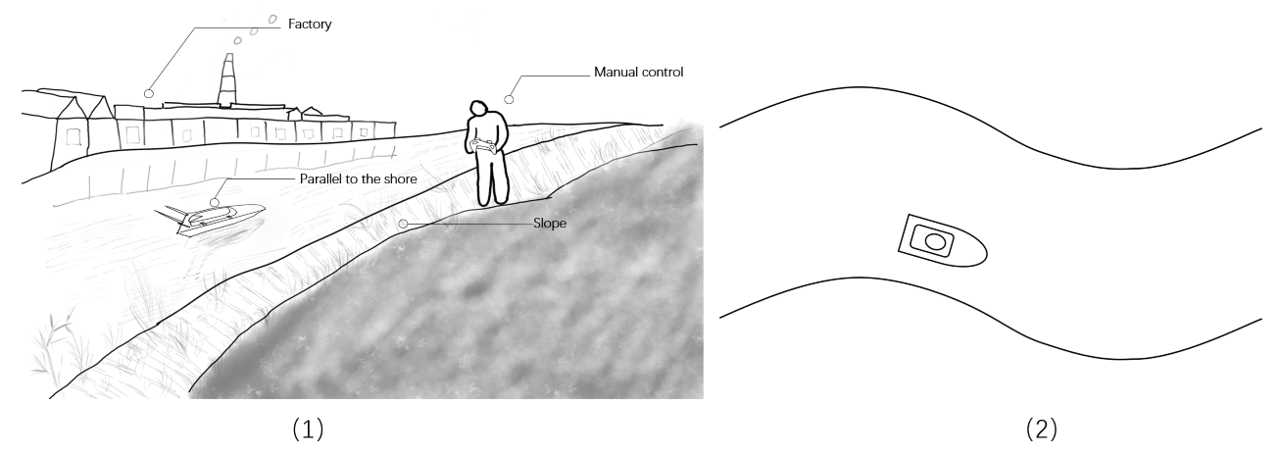}
  \caption{Operational scenario and task schematic of the underwater concealed pipe inspection: (1) Schematic representation of the riverine operation environment, including factory and slope. (2) Illustration of the USV’s parallel shoreline inspection task.} 
  \Description{It sketches the scenario two: Underwater Concealed Pipe Inspection. (1) Schematic representation of the riverine operation environment, including factory and slope. (2) Illustration of the USV’s parallel shoreline inspection task.}
  \label{Fig:S2}
\end{figure}

\paragraph{Scenario Description}The operation takes place in a narrow river channel spanning approximately 5 kilometres, with an average width of 10 meters and a water depth ranging from 1.5 to 2 meters. The western bank is adjacent to a large industrial zone suspected of being a potential pollution source, while the eastern bank borders a densely populated residential area. Both banks feature hard revetments combined with gentle earthen slopes (approximately 20 $^{\circ}$ incline), posing slip hazards for onshore personnel. Environmental challenges include turbid water with limited visibility, slow currents interspersed with shallow zones (depth < 1m in some areas), dense algae coverage exceeding 20\% of the water surface, and temporarily restricted sections marked by buoy arrays (see Figure \ref{Fig:S2} (1)). The USV equipped with a high-resolution side-scan sonar system, is tasked with conducting parallel shoreline (see Figure \ref{Fig:S2} (2)) inspections along both banks. It must maintain a precise 2-meter standoff distance from the shore while cruising at a constant low speed of approximately 0.5 m$/$s. Operators face dual responsibilities: manually controlling the USV to ensure strict parallel navigation through real-time adjustments and simultaneously analysing sonar imagery to identify suspected concealed pipes. When potential pipelines are detected, the operator must execute repetitive verification manoeuvres. This involves precisely controlling the USV to perform forward-backwards scanning passes (more than 3 repetitions) over target locations.

\paragraph{Discussion Highlights}

This scenario exemplifies the intricate challenges of human-machine collaboration in complex aquatic environments. Operators are tasked with maintaining parallel navigation along the shoreline while simultaneously interpreting sonar data to identify concealed pipes—a dual responsibility that imposes significant cognitive demands. The requirement to visually cross-reference real-time sonar imagery with direct shoreline observations creates a sensory bifurcation: operators must process high-frequency sonar (750 kHz-1.2 MHz) while monitoring the USV’s positional relationship to irregular shorelines composed of hard revetments and unstable earthen slopes. This cognitive load is further compounded by the need to navigate treacherous onshore terrain, where 20° inclines and debris-laden paths demand constant vigilance to avoid slips or falls. These physical and cognitive stressors are compounded by time-sensitive task sequences, where critical decisions must be made within constrained operational windows.
To mitigate these challenges, adaptive interface designs are essential. Eye-tracking systems integrated into the operator’s headgear equipped with AR can monitor pupil dilation as an indicator of cognitive strain. When elevated load is detected, the interface can be simplified in real-time to reduce visual clutter. Hierarchical alert systems prioritise threats through multimodal feedback. For instance, imminent collision risks (less than 1.5 metres from shore) activate haptic vibrations paired with high dB directional alarms, while high-confidence pipe detections (above 85\% confidence) prompt AR highlighting without interrupting workflow. Environmental fusion techniques enhance situational awareness. Environmental fusion techniques further enhance situational awareness. A laser projection system mounted on the USV casts visible navigation paths onto the water surface—projecting 0.3 metre wide lines visible up to 50 metres under favourable conditions. These visual guides delineate safe operational corridors and mark hazardous zones using pulsating red patterns, allowing operators to make more confident and timely decisions.

\subsubsection{Scenario Three: Post-Construction Hydrographic Survey}
\begin{figure}[h]
  \centering
 \includegraphics[width=\linewidth]{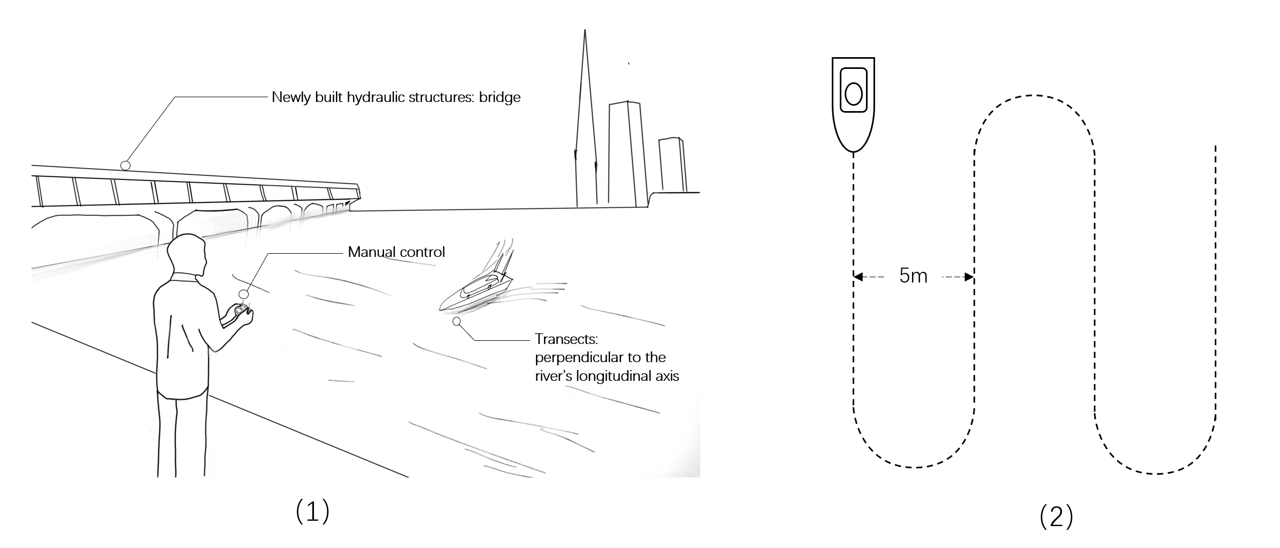}
  \caption{Operational scenario and task schematic of the post-construction hydrographic survey: (1) Manual control of perpendicular transects relative to the river’s longitudinal axis, with operators maintaining line-of-sight from the bank. (2) Example of tightly spaced (5 m) cross-sectional runs.} 
  \Description{It sketches the scenario three: Post-Construction Hydrographic Survey. (1) Manual control of perpendicular transects relative to the river’s longitudinal axis, with operators maintaining line-of-sight from the bank. (2) Example of tightly spaced (5 m) cross-sectional runs.}
  \label{Fig:S3}
\end{figure}

\paragraph{Scenario Description}This scenario unfolds in narrow, post-construction water channels adjacent to newly built hydraulic structures such as bridges, dams, or embankments. The operation aims to detect illegal dumping of construction debris, identify topographical changes, and assess potential riverbank collapses caused by the recent developments. Due to the inaccuracy (spatial deviation) and the insufficient resolution of electronic maps, USVs equipped with RTK positioning systems are operated entirely via manual control. Operators typically stand onshore and maintain a visual line-of-sight with the USV throughout the mission. Transects are planned perpendicular to the river’s longitudinal axis to enable systematic cross-sectional scanning, requiring operators to alternately steer the USV away from and towards their own position (see Figure \ref{Fig:S3} (1)). For example, in north-south flowing rivers, USVs are required an east-west route. The USV must perform a series of tightly spaced runs, commonly at 5-meter (see Figure \ref{Fig:S3} (2)), depending on the required map resolution and scale. This task made more challenging by natural river bends and narrow channel widths. One unique operational constraint involves frequent orientation reversals: when the USV’s bow faces the operator, steering directions become counterintuitive, increasing the likelihood of spatial disorientation. This results in a distinct challenge as we talk in section \ref{subsubsection:Embodied cognition}: continuously shift perspectives between operator-centric and USV-centric references.

\paragraph{Discussion Highlights}
This scenario highlights the complex interplay between spatial reasoning, motor coordination, and visual-motor alignment in human-USV interaction. Operators must translate 2D survey goals, such as uniform cross-sectional coverage, into real-time motor actions under visually ambiguous conditions. When the USV's heading opposes the operator's line of sight, directional inputs on the control interface (e.g., joysticks) do not correspond intuitively to the USV’s actual movement, a misalignment that can induce errors, especially when transitioning between approach and retreat manoeuvres in crisis moment. The cognitive demand is intensified by the absence of assistive automation. Unlike waypoint-based missions, the lack of navigational aids in this fully manual context places the entire spatial calibration burden on the operator. In this scenario, precision is critical: deviation from prescribed paths greater than 1 meter can invalidate collected data, especially when the transect spacing is tightly defined (e.g., 5 meters). Real-time decision-making is further complicated by natural obstacles such as narrow riverbanks, submerged debris, or swift currents near built structures.

To mitigate these challenges, the integration of advanced headgear equipped with AR can be instrumental. Such headgear can provide real-time visual overlays, assisting operators in maintaining correct distances from the riverbanks and ensuring that survey lines remain perpendicular to the flow. By projecting virtual guidelines and distance markers onto the operator's field of view, the headgear facilitates more accurate and efficient navigation, reducing reliance on mental calculations and minimising the risk of human error. Moreover, the headgear can be configured to provide haptic or auditory feedback when deviations from the planned survey path occur, enabling immediate corrective actions. This multimodal feedback system enhances situational awareness and allows operators to focus on the task without being overwhelmed by information.

To address spatial disorientation, it may prioritise directional consistency through multimodal feedback. AR interfaces—such as headgear —should overlay virtual arrows aligned with the USV’s bow orientation, dynamically adjusting to heading changes. One scheme to be validated is mirrored control modes, where joystick inputs are automatically inverted when the USV approaches the operator, ensuring perceived movement direction matches control inputs (e.g., pushing the joystick right steers the USV right from the operator’s viewpoint). 

\section{Discussion}
\label{section4}
Our study extends the literature on HRI by highlighting how the USV domain presents distinctive challenges for interaction design, beyond those observed in aerial or ground robotics. Prior work on unmanned aerial vehicles and autonomous driving has extensively examined issues of operator workload, trust, and automation transparency. However, the findings reported here reveal how the underactuated dynamics of USVs, combined with environmental uncertainties such as waves, currents, and unstable communication, create qualitatively different usability demands. This indicates that established HRI principles cannot be transferred wholesale to USVs, and instead require careful adaptation to the aquatic context.

Another contribution lies in advancing the understanding of operator learning and control familiarity in underexplored maritime HRI. Whereas ground and aerial systems often allow novices to gain proficiency through incremental practice in relatively predictable environments, USV operation demands embodied engagement with hydrodynamic forces and environmental disturbances that are less intuitive. Our findings highlight how operators develop coping strategies and mental models to bridge the gap between system limitations and real-world requirements. These insights not only inform interface design but also contribute to broader HRI theories of skill acquisition and embodied interaction.

Our results also underscore the importance of supporting situational awareness and decision-making under uncertainty. In contrast to environments where sensor reliability and obstacle classification are relatively stable, maritime perception is constrained by ambiguous obstacle penetrability, limited map accuracy, and communication latency. These limitations necessitate hybrid modes of autonomy and human intervention, suggesting that future research in maritime HRI should prioritise designs that make environmental uncertainty more legible and support adaptive human–automation collaboration.

It is important to acknowledge that our study has certain limitations. While our methodology combined interviews and field observations, the participant sample size and scope of environments were limited, and much of the data was self-reported. These constraints suggest caution in generalizing the findings. A more detailed discussion of these limitations is provided in Section \ref{limitation}. Nonetheless, this work establishes a foundational understanding of USV-specific usability challenges and adaptive strategies, contributing a basis for theorising and designing future maritime HRI systems.

\section{Rethinking/Design inspiration}
\label{section5}
This section reflects on insights derived from operator feedback, usability evaluations, and iterative interface design processes. The goal is to reassess prevailing assumptions and articulate key design inspirations for developing next-generation USV control systems. Six primary findings emerged from this study, each revealing implications for how control interfaces can better align with human cognition, perception, and action. These findings are organised into two overarching themes: supporting learning and control familiarity, and supporting situational awareness and decision-making.

\subsection{Supporting Learning and Control Familiarity}
Focus: Control design, usability, and embodied engagement.
Ease of use is crucial for ensuring that operators can perform their tasks with minimal frustration and maximum efficiency. A well-designed manual control interface should balance precision with ease of operation, while also considering the ergonomic and cognitive needs of the user. With the development of USV technology, continuous user feedback and iterative design improvements will be essential in creating more intuitive and effective control systems.

\textbf{Principle 1: Aligning control schemes with familiar operational habits significantly reduces the learning curve.}
Operators with prior experience in driving, aviation, or maritime control demonstrate faster adaptation when the interface logic mirrors conventional control paradigms \cite{Kirsh}. For instance, throttle-based accelerators or steering mechanisms like those in automobiles or aircraft promote immediate familiarity. Users can rely on muscle memory and prior training by embedding such metaphors directly into joystick and input device designs, including tactile cues such as detents or resistance zones, reducing hesitation and improving initial operational performance. These analogue mappings provide a scaffold that grounds novice operators and bridges the gap between physical interaction and digital command systems.

\textbf{Principle 2: Using embodied sensorimotor engagement, physical control systems reduce cognitive load.}
Unlike abstract screen-and-mouse systems (for example, dragging a cursor to simulate throttle adjustments), physical interfaces offer kinesthetic correspondence, allowing actions to mirror real-world dynamics more closely. This sensorimotor coupling enables operators to navigate, adjust, and respond with less reliance on cognitive translation, particularly in high-pressure or time-sensitive contexts \cite{article}. The physicality of the interface thus becomes a cognitive aid, transforming complex decision-making into embodied, intuitive responses that mirror real-world interactions.


\textbf{Principle 3: A dedicated and intuitive mode-switching system prevents operational errors.}
Effective human–system interaction demands a clear delineation between functional modes within the interface. When operators must navigate between tasks such as autonomous cruising, manual override, or docking, a lack of visual or structural cues can result in control conflicts or task confusion. To mitigate this, having distinct operating modes (e.g., navigation mode, manual mode) with visible indicators or buttons that lockout irrelevant controls in each mode can provide an additional layer of safety and usability. Furthermore,  differentiating between primary (mission-critical) and secondary (supplementary) controls allows users to prioritise their attention toward essential functions, easing cognitive demand and reducing the likelihood of inadvertent misoperations under stress.


\subsection{Supporting Situational Awareness and Decision-Making}
\textit{Focus: Enhancing operator insight via multimodal sensing and intelligent automation.}

\textbf{Principle 4: Allowing users to customise control settings enhances comfort and adaptability.}
Operators face diverse working conditions and personal preferences, both of which significantly impact performance and stress levels. Systems that offer customisable interface parameters, such as joystick sensitivity, feedback intensity, and button configuration, enable operators to shape their interaction environment according to individual needs. In high-stakes or unpredictable conditions, quickly tuning controls can make the difference between smooth operation and critical failure. Moreover, adaptive interfaces that respond to real-time physiological or behavioural signals, such as stress indicators or fatigue markers, may further optimise performance by aligning system responsiveness with the user’s current state. By recognising the variability of human users, such flexibility supports resilience, comfort, and sustained engagement.

\textbf{Principle 5: Shore-based perspective enhances spatial awareness and decision-making.}
The inherent superiority of shore-based perspective lies in its capacity to deliver three-dimensional spatial awareness—a critical yet underappreciated facet of USV interaction. Human operators inherently process more cues (visual cues and auditory signals) that transcend current sensor capabilities, explaining the persistent preference for shore perspectives despite technological advancements.

However, shore-based control also faces limitations, particularly in terms of constrained visual range and occluded views. To address these shortcomings, multimodal perspective integration has emerged as a promising strategy. Portable visualisation tools such as handheld monitoring panels with split-screen displays and augmented AR goggles (e.g., Microsoft HoloLens) allow operators to merge orthographic projections with first-person views. This integration enables operators to synthesise multiple perspectives, providing a more comprehensive understanding of the USV's environment and improving situational awareness. This enhanced multimodal awareness also facilitates the operator’s ability to visualise the environment in three dimensions. By combining depth perception with real-time data, operators can make more informed and predictive decisions, especially when avoiding obstacles that may not be visible from a top-down perspective. For example, AR goggles can highlight and render obstacles, which are detectable by USV cameras but not visible from the shore perspective, by using wireframe overlays to alert the operator. This AR technology not only enhances visual information but also helps visualise the experience of seasoned captains, enabling operators to improve their ability to assess ship posture and the hydrological environment.

Environmental features further contribute to enhancing depth perception and distance estimation. For instance, terrain landmarks (such as bridges and buoys) can serve as depth anchors, assisting operators in judging distance and water depth. The operator's body movements (such as head tilting) can be used for triangulation, further improving distance estimation accuracy. Atmospheric effects, such as light refraction and fog gradients, can also be utilised by the AR system to estimate distances, enhancing the operator's overall environmental perception. This fusion of multimodal information may significantly boost cognitive awareness, improving the accuracy and safety of decision-making.

In future interactive systems, operators may not merely see the USV—they will feel the water’s depth through light distortion, and sense distance through the AR landmarks. Such immersive perceptual augmentation may fundamentally reshape the operator’s relationship with the machine and environment alike.


\textbf{Principle 6: Reducing manual intervention through intelligent algorithms improves situational awareness and lowers cognitive load.}
{Reducing manual intervention through intelligent algorithms and perceptual aids improves situational awareness and reduces operator workload.}
Situational awareness in USV operations hinges on the operator's capacity to synthesise environmental data, predict hydrodynamic interactions, and anticipate system behaviours—a cognitive triad often restricted by operator experience.  Operators rely on a combination of environmental perception, system feedback, and cognitive strategies to assess real-time conditions, anticipate potential challenges and accumulate experience. Therefore, enhancing situational awareness requires an integrated approach that combines improved algorithmic support with advanced interactive tools to facilitate more intuitive and informed decision-making.

One promising avenue for improving situational awareness is the reduction of manual intervention through algorithmic optimisation. Current USV systems often require operators to make constant adjustments due to environmental uncertainties, such as tidal shifts or unpredictable currents. By refining path-planning algorithms and obstacle avoidance mechanisms, USVs can autonomously adapt to dynamic conditions, reducing the operator’s cognitive load. For instance, leveraging reinforcement learning-based models for tide compensation can help anticipate and mitigate hydrodynamic disruptions. Additionally, integrating digital twin simulations with APF architectures may further bridge the gap between human expertise and autonomous decision-making, providing real-time predictive insights.

\textbf{Principle 7: Simply increasing visual input channels (e.g., more screens, 180° views) does not enhance situational awareness unless human attentional limits are addressed.}
While industry and academia have made valuable attempts to develop multi-perspective perception tools (e.g., 180$^{\circ}$ cameras, VR displays, multiple screens), but they often overlook human attentional limitations. S-Chen pointed out even with 180$^{\circ}$ camera coverage, beginners typically fixate on less than 100$^{\circ}$ frontal sectors, due to cognitive load saturation. They exhibit strong interface-locked position dependence: most users fixate on screen-referenced waypoints rather than environmental landmarks. This contrasts with experts' hydrospatial embodiment. As S-Chen notes, "Beginners drown in multiple windows—they see everything but comprehend nothing." This highlights a key issue: simply adding more information or multiple screens does not guarantee better situational awareness. Users often find it difficult to quickly find effective information from multiple perspectives during operation, especially under dynamic hydrological conditions, which leads to some users still using a single perspective. Moreover, S-Sun’s investigations also confirm this viewpoint: simply adding more screens does not linearly enhance operators’ perceptual capacity or decision-making efficiency. To address this, their next step involves consolidating information across multiple displays into a unified 3D interface, providing an embodied spatial representation that aligns with operators’ natural modes of environmental engagement and reduces cognitive switching costs. This evolution underscores the pressing need for interfaces that simulate embodied environmental coupling.



There exists a great potential for multimodal perception tools, such as real-time visual overlays, predictive trajectory guidance, and contextual hazard alerts, which can significantly improve an operator's ability to anticipate and react to challenges. For example, AR displays could project docking assistance indicators, showing optimal approach angles and safe distances. Additionally, interactive path-planning suggestions, dynamically adjusting to environmental conditions, can help mitigate task complexity and improve mission success rates. To narrow the performance gap between beginners and experts, it can embed captain-level intuition into the interface logic through imitation learning.

In sum, this finding highlights that an effective approach to situational awareness involves both algorithmic automation to reduce manual burden and perceptual interfaces to scaffold user understanding. These two elements must work in tandem to support safe, confident, and efficient USV operations across experience levels.


\section{Limitation}
\label{limitation}
Despite the valuable insights gained from our study, there are several limitations that should be acknowledged:

\begin{itemize}
    \item Limited Participant Sample: Although we sought to ensure diversity among participants in terms of expertise and background, the limited sample size restricts the generalizability of our findings. This may have led to certain biases or missed user needs that could emerge in a larger cohort. Future work could benefit from broader sampling across geographic, institutional, and cultural contexts.
     \item Contextual and Environmental Constraints: Our study focused primarily on inland and nearshore USV operations, which may not encompass the full diversity of environments in which USVs are deployed. Including these diverse environments in future work could help uncover additional usability challenges and design considerations.
    \item Bias in Self-Reported Data: Although we conducted observational studies for cross-validation, many of the insights were derived from self-reported data, including interviews and questionnaires. While participants provided valuable feedback, self-reported data can be subject to biases such as social desirability or recall bias, potentially affecting the accuracy of the information provided.
    \item Potential for Environmental Variables: Due to the coordination challenges, the field study did not fully account for the impact of environmental factors, such as weather conditions, water currents, and visibility, which may significantly influence the usability of USVs in real-world settings. Variability in environmental conditions could affect operator performance and preferences, but was not directly addressed in this study.
\end{itemize}
These limitations suggest that future research could benefit from a larger, more diverse sample size, as well as a more comprehensive consideration of the various factors influencing USV usability across different operational contexts.

\section{Conclusion}
\label{conclusion}
In this paper, we examined the usability of USVs, through a multiple case study approach, focusing on the challenges encountered by beginners, the strategies employed by both beginners and experienced engineers, and the implications of these findings for system design. Our findings indicate that current USV systems provide a useful starting point; they are only partially viable for beginner operation in dynamic inland and offshore environments. The results highlight three persistent barriers to usability: steep learning curves for novices, reliance on embodied expertise for effective operation, and situational awareness bottlenecks caused by both technological limitations and human attentional constraints.

Through a field-observation approach, we summarized three representative use cases. Harmful algal bloom detection highlighted the tension between automated efficiency and manual precision. Underwater submerged concealed pipeline inspection revealed the conflict between sensor limitations and environmental variability. Post-construction hydrographic survey underscored the trade-off between interface complexity and the urgency of real-time operation. These use cases illustrate how operators must simultaneously manage spatial awareness, adapt to environmental variability, and make rapid decisions under uncertainty. Together, they expose limitations in current usability methodologies and highlight the necessity for more intuitive, context-aware, and effective USV interaction systems.


Building on these insights, we identified key directions for user-friendly USV interface design. First, interfaces should support learning and control familiarity by aligning with operators’ prior experience, offering embodied sensorimotor feedback, and implementing clear mode-switching mechanisms. These features reduce cognitive load, accelerate skill acquisition, and prevent operational errors. Second, interfaces should enhance situational awareness and decision-making by integrating multimodal perception, adaptive customization, and intelligent automation that reduces unnecessary manual intervention. Such approaches can bridge the gap between novices and experts, improve operational efficiency, and support safe, informed decision-making in complex maritime environments.

Overall, the findings emphasise the importance of understanding the varied mental models and learning strategies of different user groups, from beginners to experienced engineers, in order to design USV systems that are both accessible and effective. By considering the interplay between technical features, user feedback, and environmental factors, future USV designs can support smoother learning curves, reduce operational errors, and ultimately improve task performance across diverse scenarios.

In sum, current usability methods alone are insufficient. They must be re-contextualised and expanded to address the unique operational realities of USVs. By grounding this rethinking in empirical findings, our study contributes actionable insights into the design of next-generation USV interfaces. Future research should further investigate the long-term impact of analogical learning on user proficiency, as well as to explore the effectiveness of hybrid control systems in real-world environments. This research contributes to the growing body of knowledge on HRI in maritime contexts, providing actionable insights that can inform the development of more intuitive, efficient, and user-friendly unmanned vehicle systems.

\begin{acks}
The authors would like to express their sincere gratitude to:
\begin{itemize}
    \item \textbf{Mr. Kailun Liao} (Ph.D. candidate, Xi'an Jiaotong-Liverpool University) for refining the manuscript's structure and assisting with figure drawing.
    

    \item \textbf{Mr. Zehong Wu} (Ph.D. candidate, University of Science and Technology of China) for assistance in participant recruitment and valuable suggestions on the paper structure.
        
    \item \textbf{Mr. Zhuoyi Liang} (independent researcher, graduate of Tsinghua University, currently based in Zurich) for professional review of the grammatical and proofreading of the manuscript.

\end{itemize}
\end{acks}

\bibliographystyle{ACM-Reference-Format}
\bibliography{samples/sample-base}

\appendix

\newpage
\appendix
\label{appendix}
\section{Interview Questions for Control Engineers and Solution Architects}
\begin{enumerate}
    \item Socio-Demographic Question
        \begin{itemize}
            \item What is your age?
            \item How many years of experience do you have in USV?
            \item What types of tasks are you mainly responsible for USV in the company?
        \end{itemize}
    \item Current System Evaluation
        \begin{itemize}
            \item What are the main functions of the current USV operation system?
                \begin{itemize}
                    \item How would you evaluate its performance?
                \end{itemize}
            \item What are the most common problems you encounter during USV operations?
            \item Are there any features or tools you wish the current system had?
            \item What interaction methods are currently used for USV operations? 
                \begin{itemize}
                    \item What are their advantages and disadvantages?
                    \item What other interaction methods would you like to see added?
                \end{itemize}
        \end{itemize}
    \item  Common Issues and Guidance Strategies (For Control Engineers)
        \begin{itemize}
            \item What are the most common issues you encounter when training beginners on USVs?
            \item What do you think are the main sources of these difficulties (e.g., interface design, operation logic, complexity, user background)?
            \item What strategies do you use to help beginners better understand and operate USVs?
            \item Have you used specific tools or training materials to assist beginners?
            \item What strategies or methods do you find most effective, and why?
            \item How do you determine if a new user has fully mastered USV control? (e.g., what tasks can they independently complete to show competence?)
        \end{itemize}
    \item Interaction Design Considerations (For Solution Architects)
        \begin{itemize}
            \item How do you consider the needs and challenges of beginners when designing the interface for USVs?
            \item What are your assumptions behind these design choices?
            \item In your opinion, what aspects of the current interface help beginners learn and use the USV effectively? What criteria do you use for this evaluation?
            \item What areas need improvement to better support beginners?
            \item Have you used specific methods or tools to test and optimise the user experience for beginners?
        \end{itemize}
    \item Task Scenarios
        \begin{itemize}
            \item Can you describe a typical task or the most frequent task involving USV operations?
            \item What is the purpose of the task? Describe the specific USV (appearance, performance).
            \item What is the environment like? (location, season, time of day, weather, any constraints?)
            \item What steps are involved in the task? Please describe the sequence.
            \item What are the criteria for task completion?
            \item Which steps are the most time-consuming or error-prone?
            \item What improvements could help simplify these steps or reduce errors?
        \end{itemize}
    \item Interaction Suggestions
        \begin{itemize}
            \item Do you have experience using AR or VR devices? If so, how do you find their interaction methods?
            \item What are your initial thoughts on using eye-tracking and gesture control for USVs?
            \item In the task scenarios you've described, which operations could be handled through eye-tracking and gesture control? 
            \item What gestures seem realistic to implement, and why?
            \item Can you think of any specific eye-tracking gestures or controls that would make USV operations more intuitive for beginners?
        \end{itemize}
\end{enumerate}

\section{Interview Questions for Users}
\begin{enumerate}
\item Interviewee Background
   \begin{itemize}
    \item Interviewee's name, position, and affiliated organisation.
    \item Their experience with USV operations (duration, frequency).
    \end{itemize}

\item Operational Experience
\begin{itemize}
    \item Describe specific tasks involving USV operation (e.g., data collection, monitoring).
    \item What are the main challenges and considerations during operation?
\end{itemize}

\item Training and Support
\begin{itemize}
    \item What kind of training did you receive before operating USVs?
    \item Which training content was most helpful in actual operations?
\end{itemize}

\item System Feedback
\begin{itemize}
    \item What shortcomings do you see in the current USV interaction system?
    \item What features or functionalities would you like to see improved (UI, operation process, etc.)?
\end{itemize}

\item Evaluation Standards
\begin{itemize}
    \item For the typical use cases you've mentioned, what evaluation criteria do you consider important (e.g., operational efficiency, data accuracy, user satisfaction)?
    \item How do you assess the performance of USVs in actual operations?
\end{itemize}

\item Procurement Decisions (For Research Institutes or Government Agencies)
\begin{itemize}
    \item What is the primary purpose of procuring USVs (e.g., data collection, environmental monitoring)?
    \item What factors are most important when selecting a USV?
\end{itemize}

\item Future Outlook
\begin{itemize}
    \item What do you see as the future development direction for USV technology?
    \item What role do you expect USVs to play in scientific research or environmental monitoring?
\end{itemize}

\end{enumerate}

\end{document}